\documentclass[aps,preprint,amsmath,amssymb]{revtex4}
\usepackage{graphicx}
\usepackage[maxfloats=32]{morefloats}
\begin{document}

\title{Probe of anomalous quartic $WWZ\gamma$ couplings in photon-photon collisions}

\author{A. Senol}
\email[]{senol_a@ibu.edu.tr} \affiliation{Department of Physics,
Abant Izzet Baysal University, 14280, Bolu, Turkey}

\author{M. K\"{o}ksal}
\email[]{mkoksal@cumhuriyet.edu.tr} \affiliation{Department of
Physics, Cumhuriyet University, 58140, Sivas, Turkey}

\begin{abstract}
In this paper, we examine the potentials of the processes $\gamma
\gamma\rightarrow W^{+} W^{-}Z $  and $e^{+}e^{-} \rightarrow
e^{+}\gamma^{*} \gamma^{*} e^{-} \rightarrow e^{+} W^{+} W^{-} Z
e^{-}$ at the CLIC with $\sqrt{s}=0.5,1.5$ and $3$ TeV to
investigate anomalous quartic $WWZ\gamma$ couplings by two different
CP-violating and CP-conserving effective Lagrangians. We find $95\%$
confidence level sensitivities on the anomalous coupling parameters
at the three CLIC energies and various integrated luminosities. The
best sensitivities obtained from the process $\gamma
\gamma\rightarrow W^{+} W^{-}Z $ on the anomalous
$\frac{k_{0}^{W}}{\Lambda^{2}}$, $\frac{k_{c}^{W}}{\Lambda^{2}}$ and
$\frac{k_{2}^{m}}{\Lambda^{2}}$ couplings defined by CP-conserving
effective Lagrangians are $[-1.73;\, 1.73]\times 10^{-7}$
GeV$^{-2}$, $[-2.44;\, 2.44]\times 10^{-7}$ and $[-1.89; \,
1.89]\times 10^{-7}$ GeV$^{-2}$, while $\frac{a_{n}}{\Lambda^{2}}$
coupling determined by CP-violating effective Lagrangians is
obtained as $[-1.74;\, 1.74]\times 10^{-7}$ GeV$^{-2}$. In addition,
the best sensitivities derived on $\frac{k_{0}^{W}}{\Lambda^{2}}$,
$\frac{k_{c}^{W}}{\Lambda^{2}}$ and $\frac{k_{2}^{m}}{\Lambda^{2}}$
and $\frac{a_{n}}{\Lambda^{2}}$ from the process $e^{+}e^{-}
\rightarrow e^{+}\gamma^{*} \gamma^{*} e^{-} \rightarrow e^{+} W^{+}
W^{-} Z e^{-}$ are obtained as $[-1.09;\, 1.09]\times 10^{-6}$
GeV$^{-2}$, $[-1.54;\, 1.54]\times 10^{-6}$ GeV$^{-2}$, $[-1.18;\,
1.18]\times 10^{-6}$ and $[-1.04;\, 1.04]\times 10^{-6}$ GeV$^{-2}$,
respectively.
\end{abstract}

\maketitle

\section{Introduction}

Gauge boson self-couplings are completely defined by the non-abelian
$SU(2)\times U(1)$ gauge symmetry of the Standard Model (SM), thus
direct search for these couplings are extremely significant in
understanding the gauge structure of the SM. However, the possible
deviation from the SM predictions of gauge boson self-couplings
would be a sign for the presence of new physics beyond the SM. Probe
of the new physics in a model independent way by means of the
effective Lagrangian approach is often a common way. In this
approach, anomalous quartic gauge boson couplings are described by
means of high-dimensional effective operators and they do not cause
anomalous trilinear gauge boson couplings. Therefore, anomalous
quartic gauge boson couplings can be independently investigated from
any trilinear gauge boson couplings.

In the literature, the anomalous quartic $WWZ\gamma$ couplings are
usually investigated by two different dimension 6 effective
Lagrangians that keep custodial $SU(2)_{c}$ symmetry and local
$U(1)_{QED}$ symmetry. The first is CP-violating effective
Lagrangian. It is defined by \cite{lag1}
\begin{eqnarray}
\textit{L}_{n}=\frac{i \pi \alpha}{4\Lambda^{2}} a_{n} \epsilon_{ijk} W_{\mu \alpha}^{(i)}W_{\nu}^{(j)} W^{(k)\alpha} F^{\mu\nu}
\end{eqnarray}
where $F^{\mu \nu}$ is the tensor for electromagnetic field
strength, $\alpha=\frac{e^{2}}{4\pi}$ is the fine structure
constant, $a_{n}$ is the dimensionless anomalous quartic coupling
constant and $\Lambda$ is represented the energy scale of new
physics. The anomalous $W W Z \gamma $ vertex function obtained from
effective Lagrangian in Eq. $1$ is given in Appendix.

Secondly, we apply the formalism of Ref. \cite{lhc} to examine
CP-conserving effective Lagrangian. As can be seen from Eq. 5 in
Ref. \cite{lhc}, there are fourteen effective photonic operators
related to the anomalous quartic gauge couplings. These operators
are identified by fourteen independent couplings
$k_{0,c}^{w,b,m},k_{1,2,3}^{w,m}$ and $k_{1,2}^{b}$. However, the
effective interactions in these operators can be expressed in terms
of independent Lorentz structures. For example, the $WW\gamma\gamma$
and $ZZ\gamma\gamma$ interactions can be parameterized in terms of
four independent Lorentz structures,
\begin{eqnarray}
\textit{W}_{0}^{\gamma}&=&\frac{-e^{2}g^{2}}{2}F_{\mu \nu}F^{\mu
\nu} W^{+ \alpha} W_{\alpha}^{-},
\\
\textit{W}_{c}^{\gamma}&=&\frac{-e^{2}g^{2}}{4}F_{\mu \nu}F^{\mu
\alpha} (W^{+ \nu} W_{\alpha}^{-}+W^{- \nu} W_{\alpha}^{+}),
\end{eqnarray}
\begin{eqnarray}
\textit{Z}_{0}^{\gamma}&=&\frac{-e^{2}g^{2}}{4
\textmd{cos}^{2}\,\theta_{W}}F_{\mu \nu}F^{\mu \nu} Z^{\alpha}
Z_{\alpha},
\\
\textit{Z}_{c}^{\gamma}&=&\frac{-e^{2}g^{2}}{4
\textmd{cos}^{2}\,\theta_{W}}F_{\mu \nu}F^{\mu \alpha} Z^{\nu}
Z_{\alpha}.
\end{eqnarray}
Also, among them two are related to $ZZZ\gamma$ operators:
\begin{eqnarray}
\textit{Z}_{0}^{Z}&=&\frac{-e^{2}g^{2}}{2
\textmd{cos}^{2}\,\theta_{W}}F_{\mu \nu}Z^{\mu \nu} Z^{\alpha}
Z_{\alpha},
\\
\textit{Z}_{c}^{Z}&=&\frac{-e^{2}g^{2}}{2
\textmd{cos}^{2}\,\theta_{W}}F_{\mu \nu}Z^{\mu \alpha} Z^{\nu}
Z_{\alpha}.
\end{eqnarray}
The remaining $WWZ\gamma$ interactions are given as follows
\begin{eqnarray}
\textit{W}_{0}^{Z}&=&-e^{2}g^{2} F_{\mu \nu}Z^{\mu \nu} W^{+ \alpha}
W_{\alpha}^{-},
\\
\textit{W}_{c}^{Z}&=&-\frac{e^{2}g^{2}}{2}F_{\mu \nu}Z^{\mu \alpha}
(W^{+ \nu} W_{\alpha}^{-}+W^{- \nu} W_{\alpha}^{+})
\\
\textit{W}_{1}^{Z}&=&-\frac{e g_{z}g^{2}}{2}F^{\mu \nu} (W_{\mu
\nu}^{+}W_{\alpha}^{-} Z^{\alpha}+W_{\mu \nu}^{-}W_{\alpha}^{+}
Z^{\alpha})
\\
\textit{W}_{2}^{Z}&=&-\frac{e g_{z}g^{2}}{2}F^{\mu \nu} (W_{\mu
\alpha}^{+} W^{-\alpha} Z_{\nu}+W_{\mu \alpha}^{-}W^{+ \alpha}
Z_{\nu})
\\
\textit{W}_{3}^{Z}&=&-\frac{e g_{z}g^{2}}{2}F^{\mu \nu} (W_{\mu
\alpha}^{+} W^{-}_{\nu} Z^{\alpha}+W_{\mu \alpha}^{-}W_{\nu}^{+}
Z^{\alpha})
\end{eqnarray}
with $g=e/ s_W$, $g_{z}=e/ s_W c_W$ and
$V_{\mu\nu}=\partial_{\mu}V_{\nu}-\partial_{\nu}V_{\mu}$ where
$s_W=\textmd{sin}\,\theta_{W}$, $c_W=\textmd{cos}\,\theta_{W}$ and
$V=W^{\pm},Z$. The anomalous vertex functions obtained through the
CP-conserving anomalous $W W Z \gamma$ interactions in Eqs.
($8$)-($12$) are given in Appendix.

Therefore, the fourteen effective photonic operators related to the
anomalous quartic gauge couplings can be appropriately rewritten in
terms of the above independent Lorentz structures
\begin{eqnarray}
\textit{L}=&&\frac{k_{0}^{\gamma}}{\Lambda^2}(\textit{Z}_{0}^{\gamma}+\textit{W}_{0}^{\gamma})+\frac{k_{c}^{\gamma}}{\Lambda^2}(\textit{Z}_{c}^{\gamma}
+\textit{W}_{c}^{\gamma})+\frac{k_{1}^{\gamma}}{\Lambda^2}\textit{Z}_{0}^{\gamma} \nonumber \\
&&+\frac{k_{23}^{\gamma}}{\Lambda^2}\textit{Z}_{c}^{\gamma}+\frac{k_{0}^{Z}}{\Lambda^2}\textit{Z}_{0}^{Z}+\frac{k_{c}^{Z}}{\Lambda^2}\textit{Z}_{c}^{Z}+\sum_{i}\frac{k_{i}^{W}}{\Lambda^2}\textit{W}_{i}^{Z}
\end{eqnarray}
where the coefficients that parametrise the strength of the
anomalous quartic gauge couplings are expressed as
\begin{eqnarray}
k_{j}^{\gamma}=k_{j}^{w}+k_{j}^{b}+k_{j}^{m}\,\,\,\,\,\,\,\,\,\,\,\,\,\,\,\,(j=0,c,1)
\end{eqnarray}
\begin{eqnarray}
k_{23}^{\gamma}=k_{2}^{w}+k_{2}^{b}+k_{2}^{m}+k_{3}^{w}+k_{3}^{m}
\end{eqnarray}
\begin{eqnarray}
k_{0}^{Z}=\frac{c_W}{s_W}(k_{0}^{w}+k_{1}^{w})-\frac{s_W}{c_W}(k_{0}^{b}+k_{1}^{b})+c_{zw}(k_{0}^{m}+k_{1}^{m}),
\end{eqnarray}
\begin{eqnarray}
k_{c}^{Z}=\frac{c_W}{s_W}(k_{c}^{w}+k_{2}^{w}+k_{3}^{w})-\frac{s_W}{c_W}(k_{c}^{b}+k_{2}^{b})+c_{zw}(k_{c}^{m}+k_{2}^{m}+k_{3}^{m}),
\end{eqnarray}
\begin{eqnarray}
k_{0}^{W}=\frac{c_W}{s_W}k_{0}^{w}-\frac{s_W}{c_W}k_{0}^{b}+c_{zw}k_{0}^{m},
\end{eqnarray}
\begin{eqnarray}
k_{c}^{W}=\frac{c_W}{s_W}k_{c}^{w}-\frac{s_W}{c_W}k_{c}^{b}+c_{zw}k_{c}^{m},
\end{eqnarray}
\begin{eqnarray}
k_{j}^{W}=k_{j}^{w}+\frac{1}{2}k_{j}^{m}\,\,\,\,\,\,\,\,\,\,\,\,\,\,\,\,(j=1,2,3).
\end{eqnarray}
where $c_{zw}=(c_W^2-s_W^2)/(2c_Ws_W)$.

For this study, we take care of the five coefficients
$k_{i}^{W}$($i=0,c,1,2,3$) defined in Eqs. ($18$)-($20$)
corresponding to the $WWZ\gamma$ vertex. However, these parameters
are correlated with those coupling constants that describe $WW\gamma
\gamma, ZZ\gamma \gamma$ and $ZZZ \gamma$ couplings \cite{lhc}.
Thus, the anomalous $WWZ\gamma$ coupling should be dissociated from
the other anomalous quartic couplings to obtain the only
non-vanishing $WWZ\gamma$ vertex. For the non-vanishing of the only
$WWZ\gamma$ vertex, we can apply additional restrictions on
$k_{i}^{j}$ parameters. One of the possible restrictions, proposed
in \cite{lag3}, to verify this is to set $k_{2}^{m}=-k_{3}^{m}$ and
other parameters($k_{0,c}^{w,b,m},k_{1,2,3}^{w}, k_1^m$ and
$k_{1,2}^{b}$) to zero. As a result of this choice, Eq. ($13$)
reduces to only non-vanishing $WWZ\gamma$ couplings as follows
\begin{eqnarray}
\textit{L}_{eff}=\frac{k_{2}^{m}}{2\Lambda^{2}}(W_{2}^{Z}-W_{3}^{Z}).
\end{eqnarray}

The current experimental sensitivities on $a_{n}/\Lambda^{2}$
parameter derived from CP-violating effective Lagrangian through the
process $e^{+}e^{-}\rightarrow W^{+}W^{-} \gamma$ at the LEP are
obtained by L3, OPAL and DELPHI collaborations. These are

\begin{eqnarray}
\textit{L3}:-0.14\,  \textmd{GeV}^{-2}<\frac{a_{n}}{\Lambda^{2}}<0.13\,\textmd{GeV}^{-2},
\end{eqnarray}
\begin{eqnarray}
\textit{OPAL}:-0.16\,  \textmd{GeV}^{-2}<\frac{a_{n}}{\Lambda^{2}}<0.15\,  \textmd{GeV}^{-2},
\end{eqnarray}
\begin{eqnarray}
\textit{DELPHI}:-0.18\,  \textmd{GeV}^{-2}<\frac{a_{n}}{\Lambda^{2}}<0.14\,  \textmd{GeV}^{-2}
\end{eqnarray}
at $95\%$ confidence level \cite{lep1,lep2,lep3}.

Besides, the CERN LHC provides current experimental sensitivities on
only $\frac{k_{0}^{W}}{\Lambda^{2}}$ and
$\frac{k_{c}^{W}}{\Lambda^{2}}$ couplings given in Eqs.
($18$)-($19$) which are related to the anomalous quartic $WWZ\gamma$
couplings within CP-conserving effective Lagrangians \cite{sınır}.
The results obtained for these couplings at $95\%$ C. L. through the
process $q\overline{q}'\rightarrow W
(\rightarrow\ell\nu)Z(\rightarrow jj) \gamma$ at $\sqrt{s}=8$ TeV
with an integrated luminosity of $19.3$ fb$^{-1}$ are given as
follows
\begin{eqnarray}
-1.2\times 10^{-5}  \textmd{GeV}^{-2}<\frac{k_{0}^{W}}{\Lambda^{2}}<1\times 10^{-5} \textmd{GeV}^{-2}
\end{eqnarray}
and
\begin{eqnarray}
-1.8\times 10^{-5}  \textmd{GeV}^{-2}<\frac{k_{c}^{W}}{\Lambda^{2}}<1.7\times 10^{-5} \textmd{GeV}^{-2}.
\end{eqnarray}

There have been many studies for anomalous quartic $WWZ\gamma$
couplings at linear and hadron colliders. The linear $e^{+}e^{-}$
colliders and their operating modes of $e \gamma$ and $\gamma\gamma$
have been investigated through the processes $e^{+}e^{-}\rightarrow
W^{+}W^{-}Z,W^{+}W^{-}\gamma$ \cite{lin,lag3,linb,linc,lind,line},
$e^{+}e^{-} \rightarrow e^{+}\gamma^{*} e^{-} \rightarrow e^{+}
W^{-} Z \nu_{e}$ \cite{mur}, $e \gamma\rightarrow W^{+}W^{-}e,
\nu_{e}W^{-}Z$ \cite{lag1,lin2} and $\gamma\gamma\rightarrow
W^{+}W^{-}Z$ \cite{lin3,lin4}. In addition, a detailed analysis of
anomalous $WWZ\gamma$ couplings at the LHC have been studied via the
processes $pp\rightarrow$ $W Z\gamma$ \cite{lhc,lhc1} and
$pp\rightarrow p\gamma^{*} p\rightarrow p W Z q X$ \cite{mur1}. The
photonic quartic $WW\gamma\gamma$ and $ZZ\gamma\gamma$ couplings are
examined in photon-photon reactions, i.e. $pp\to p\gamma^*\gamma^*
p\to p W^+W^- p$
\cite{Pierzchala:2008xc,Chapon:2009hh,deFavereaudeJeneret:2009db}
for $WW\gamma\gamma$ couplings and $pp\to p\gamma^*\gamma^* p\to p Z
Z p$ \cite{Chapon:2009hh,Gupta:2011be}.

The LHC is anticipated to answer some of the unsolved questions of
particle physics. However, it may not provide high precision
measurements due to the remnants remaining after the collision of
the proton beams. A linear $e^{+}e^{-}$ collider with high
luminosity and energy is the best option to complement and to extend
the LHC physics program. The CLIC is one of the most popular linear
colliders, planned to carry out $e^{+}e^{-}$ collisions at energies
from $0.5$ TeV to $3$ TeV \cite{clic}. To have its high luminosity
and energy is quite important with regards to new physics research
beyond the SM. Since the anomalous quartic $WWZ\gamma$ couplings
described through CP-violating and CP-conserving effective
Lagrangians have dimension-6, they have very strong energy
dependences. Thus, the anomalous cross section containing the
$WWZ\gamma$ vertex has a higher energy than the SM cross section. In
addition, the future linear collider will possibly generate a final
state with three or more massive gauge bosons. Hence, it will have a
great potential to examine anomalous quartic gauge boson couplings.

Another possibility expected for the linear colliders is to operate
this machine as $\gamma \gamma$ and $\gamma e$ colliders. This can
be performed by converting the incoming leptons into intense beams
of high-energy photons \cite{las1,las2}. On the other hand,
$\gamma^{*} \gamma^{*}$ and $\gamma^{*} e$ processes at the linear
colliders arise from quasi-real photon emitted from the incoming
$e^{+}$ or $e^{-}$ beams. Hence, $\gamma^{*} \gamma^{*}$ and
$\gamma^{*} e$ processes are more realistic than $\gamma \gamma$ and
$\gamma e$ processes. The photons in these processes are defined by
the Equivalent Photon Approximation (EPA) \cite{Brodsky:1971ud,
Terazawa:1973tb, es1,es2,es3}. In the EPA, the quasi-real photons
are scattered at very small angles from the beam pipe, so they have
low virtuality. For this reason, they are supposed to be almost
real. Moreover, the EPA has a lot of advantages: First, it provides
the skill to reach crude numerical predictions via simple formulae.
In addition, it may principally ease the experimental analysis
because it enables one to achieve directly a rough cross section for
$\gamma^{*} \gamma^{*}\rightarrow X$ process via the examination of
the main process $e^{+}e^{-}\rightarrow e^{+} X e^{-}$. Here, $X$
represents objects produced in the final state. The production of
high mass objects is specially interesting at the linear colliders.
Furthermore, the production rate of massive objects is limited by
the photon luminosity at high invariant mass.

In conclusion, these processes have a very clean experimental
environment, since they have no interference with weak and strong
interactions. Up to now, the photon-induced processes for the new
physics searches were investigated through the EPA at the LEP,
Tevatron, LHC and CLIC in literature
\cite{a1,a2,a3,a4,a5,a6,a7,a8,a9,a10,a11,a12,a13,a14,a16,a17,a18,a19,a20,a21,a22,a23,a24,a25,a26,a27,a28,a29,a30,murc1,murc2,murc3,murc4,murc5}.

\section{CROSS SECTIONS AND NUMERICAL ANALYSIS }

All numerical calculations in this study were evaluated using the
computer package CalcHEP \cite{calc} by embedding the anomalous
$WWZ\gamma$ interaction vertices defined through CP-violating
[Eq.($1$)] and CP-conserving [Eqs. ($8$)-($12$)] effective
operators. The total cross sections for two processes $\gamma
\gamma\rightarrow W^{+} W^{-}Z$ and $e^{+}e^{-}\rightarrow
e^{+}\gamma^{*} \gamma^{*} e^{-} \rightarrow e^{+} W^{+} W^{-} Z
e^{-}$  in terms of $k_i^{W}$ ($i=0,c$) couplings can be given by
\begin{eqnarray}\label{tcs}
\sigma_{tot}=\sigma_{SM}+\sum_i\frac{k_i^{W}}{\Lambda^2}\sigma_{int}^i+\sum_{i,j}\frac{k_i^{W}k_j^{W}}{\Lambda^4}\sigma_{ano}^{ij}.
\end{eqnarray}
In addition, the total cross sections containing $k_2^{m}$ couplings are obtained as follows
\begin{eqnarray}\label{tcs}
\sigma_{tot}=\sigma_{SM}+\frac{k_2^{m}}{\Lambda^2}\sigma_{int}+\frac{(k_2^{m})^{2}}{\Lambda^4}\sigma_{ano}.
\end{eqnarray}
Finally, the total cross sections including $a_n$ couplings can be written by
\begin{eqnarray}\label{tcs}
\sigma_{tot}=\sigma_{SM}+\frac{a_n^{2}}{\Lambda^4}\sigma_{ano}
\end{eqnarray}
where $\sigma_{SM}$ is the SM cross section, $\sigma_{int}$ is the
interference terms between SM and the anomalous contribution, and
$\sigma_{ano}$ is the pure anomalous contribution. The interference
terms in total cross sections given in Eqs. $(27)$-$(28)$ related to
CP-conserving effective Lagrangians are negligibly small compared to
pure anomalous terms. Nevertheless, we took into account the effect
of all interference term in the numerical calculations. However, the
total cross section depends only on the quadratic function of $a_n$
anomalous coupling defined by CP-violating effective Lagrangians,
since anomalous coupling $a_n$ does not interfere with the SM
amplitude.

\begin{figure}
\includegraphics[width=0.8\columnwidth] {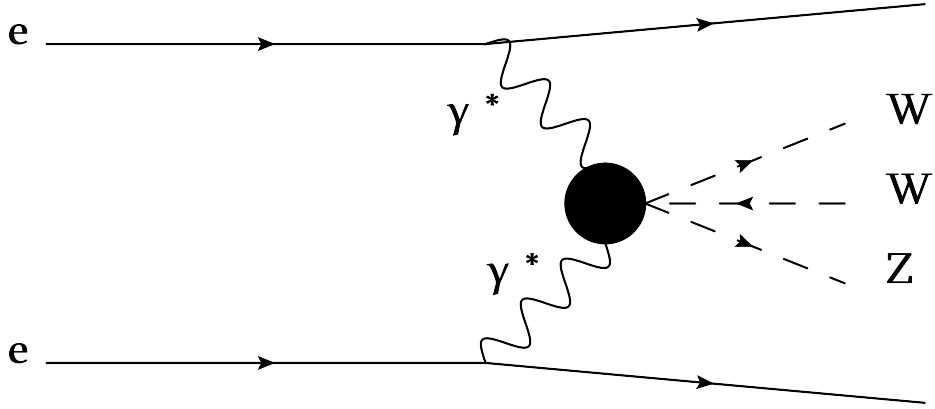}
\caption{Schematic diagram for the process $e^{+}e^{-}\rightarrow
e^{+}\gamma^{*} \gamma^{*}e^{-}\rightarrow e^{+}W^{+} W^{-} Z e^{-}$
at the CLIC. \label{fig1}}
\end{figure}

\begin{figure}
\includegraphics[width=0.8\columnwidth]{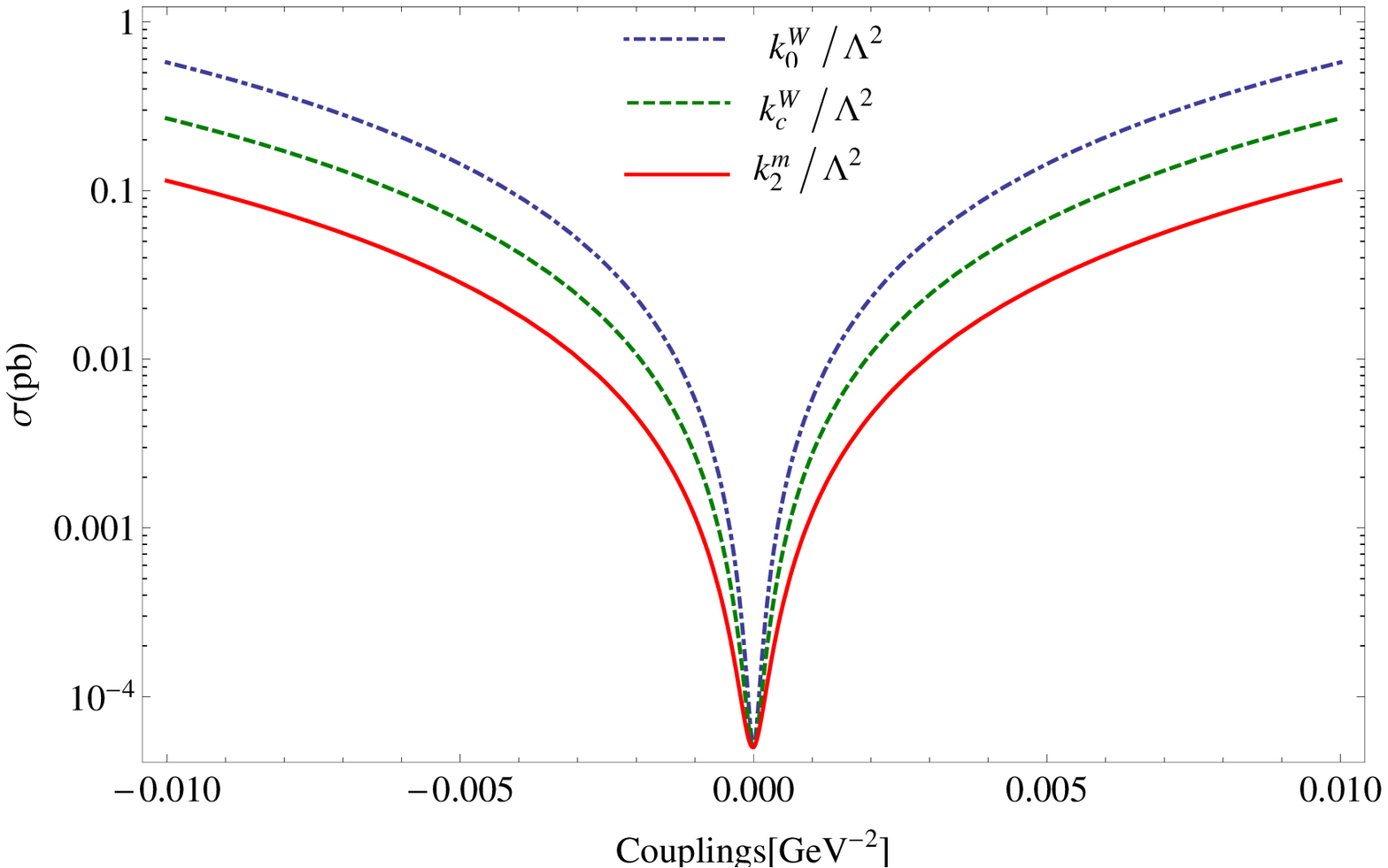}
\caption{The total cross sections as function of anomalous
$\frac{k_{0}^{W}}{\Lambda^{2}}$, $\frac{k_{c}^{W}}{\Lambda^{2}}$ and
$\frac{k_{2}^{m}}{\Lambda^{2}}$ couplings for the  process
$e^{+}e^{-}\rightarrow e^{+}\gamma^{*} \gamma^{*}e^{-}\rightarrow
e^{+}W^{+} W^{-} Z e^{-}$ at the CLIC with $\sqrt{s}=0.5$ TeV.
\label{fig2}}
\end{figure}

\begin{figure}
\includegraphics[width=0.8\columnwidth]{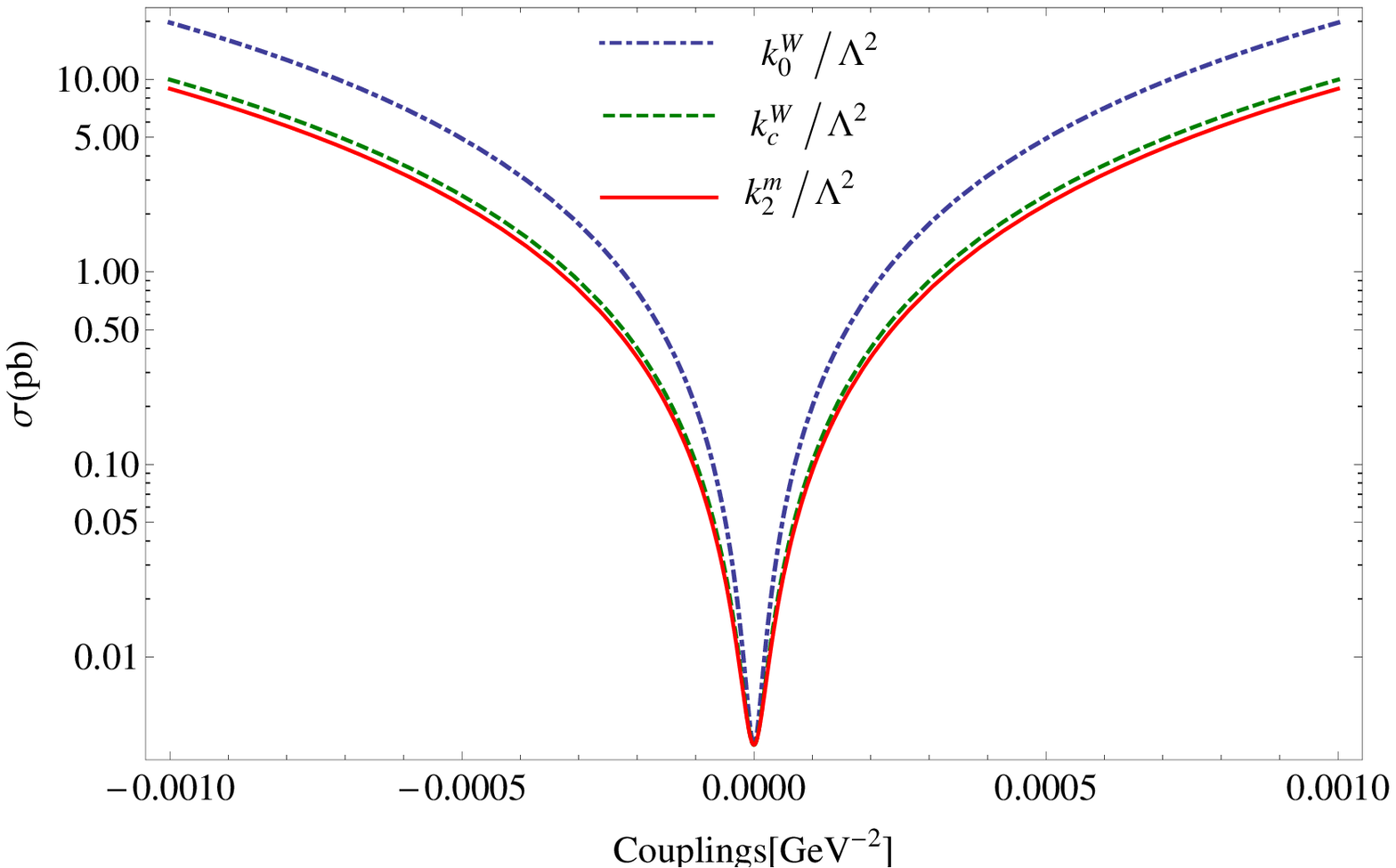}
\caption{The same as Fig. 2 but for $\sqrt{s}=1.5$ TeV.
\label{fig3}}
\end{figure}

\begin{figure}
\includegraphics[width=0.8\columnwidth]{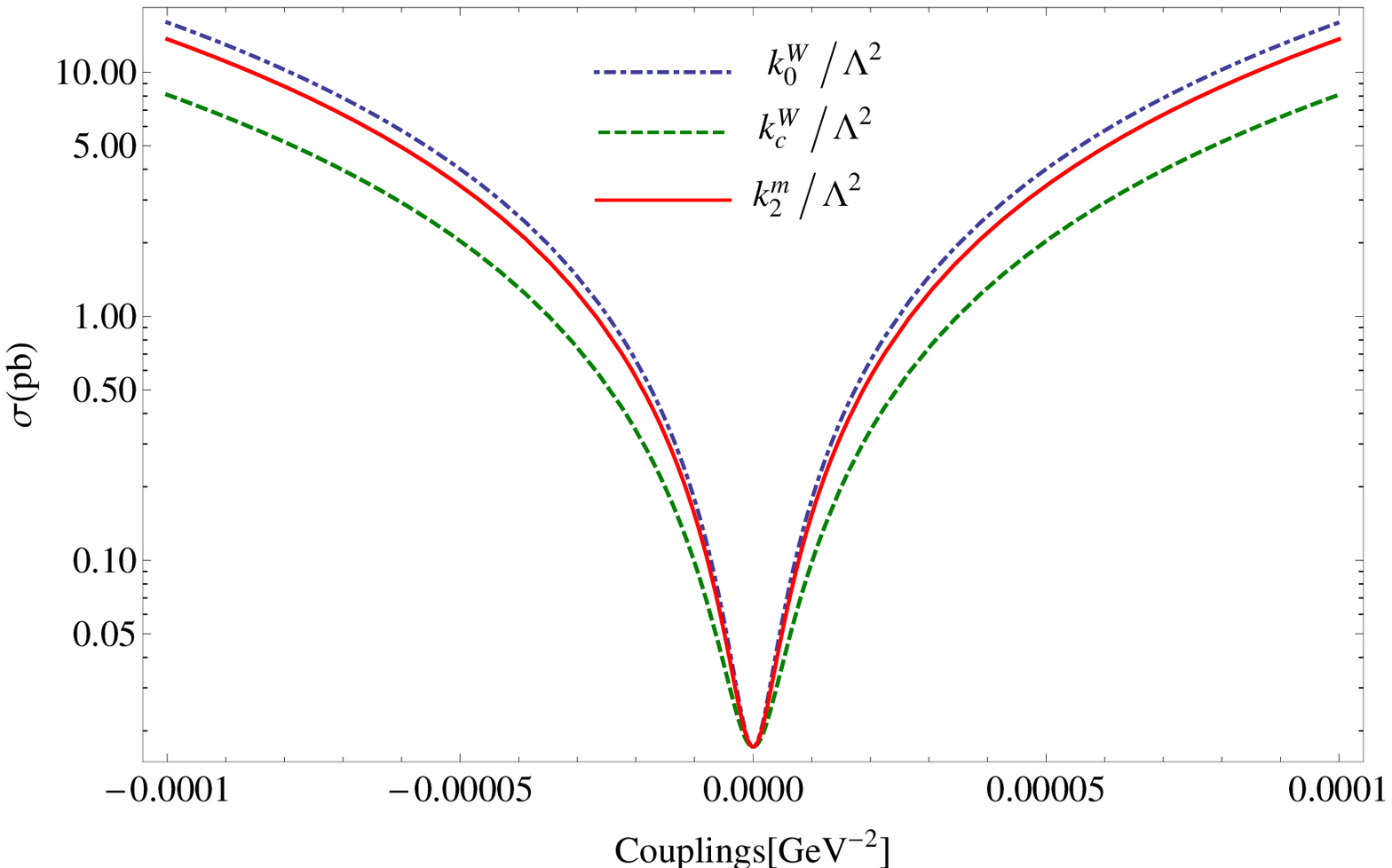}
\caption{The same as Fig. 2 but for $\sqrt{s}=3$ TeV.
\label{fig4}}
\end{figure}

\begin{figure}
\includegraphics [width=0.8\columnwidth] {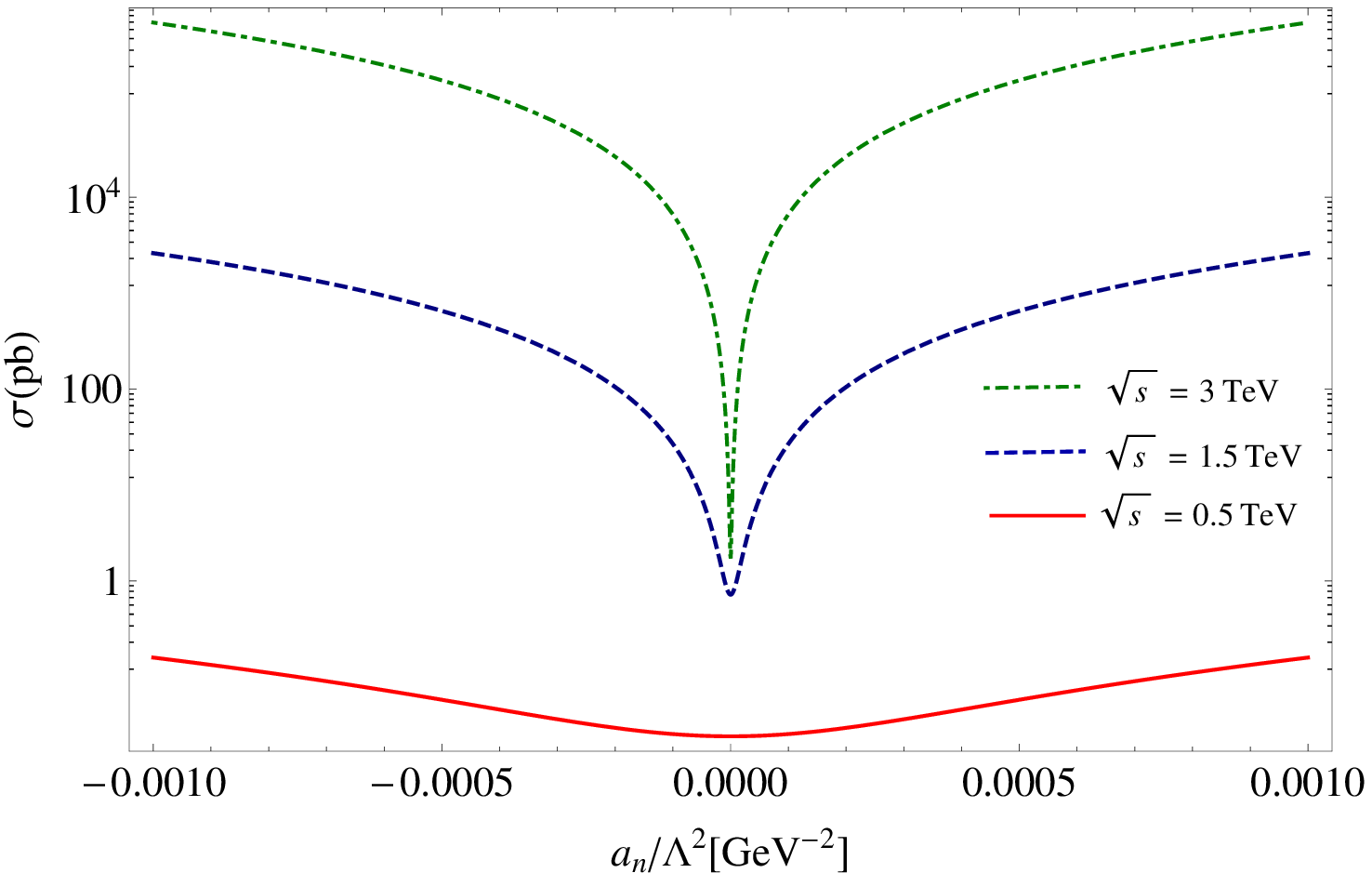}
\caption{The total cross sections as function of anomalous
$\frac{a_{n}}{\Lambda^{2}}$ coupling for the process
$e^{+}e^{-}\rightarrow e^{+}\gamma^{*} \gamma^{*}e^{-}\rightarrow
e^{+}W^{+} W^{-} Z e^{-}$ at the CLIC with $\sqrt{s}=0.5, 1.5$ and
$3$ TeV. \label{fig5}}
\end{figure}

The quasi-real photons emitted from both lepton beams collide with
each other, and the process $\gamma^{*} \gamma^{*}\rightarrow W^{+}
W^{-}Z$ is generated. The process $\gamma^{*} \gamma^{*}\rightarrow
W^{+} W^{-}Z$ participates as a subprocess in the main process
$e^{+}e^{-}\rightarrow e^{+}\gamma^{*} \gamma^{*} e^{-} \rightarrow
e^{+} W^{+} W^{-} Z e^{-}$. A schematic diagram representing the
main process is given in Fig. \ref{fig1}. When calculating the total
cross sections for this process, we used the equivalent photon
spectrum described by the EPA which is embedded in CalcHEP. The
total cross sections of the process $e^{+}e^{-}\rightarrow
e^{+}\gamma^{*} \gamma^{*} e^{-} \rightarrow e^{+} W^{+} W^{-} Z
e^{-}$ as functions of anomalous $\frac{k_{0}^{W}}{\Lambda^{2}}$,
$\frac{k_{c}^{W}}{\Lambda^{2}}$, $\frac{k_{2}^{m}}{\Lambda^{2}}$ at
$\sqrt s$=0.5, 1.5 and 3 TeV are shown in Figs.
\ref{fig2}-\ref{fig4}, respectively. Dependence of the
$e^{+}e^{-}\rightarrow e^{+}\gamma^{*} \gamma^{*} e^{-} \rightarrow
e^{+} W^{+} W^{-} Z e^{-}$ cross section on the anomalous
$\frac{a_{n}}{\Lambda^{2}}$ couplings at the same three
center-of-mass energies are given in Fig. \ref{fig5}. Here, we
assume that only one of the anomalous couplings deviate from the SM
at any given time. We can see from Figs. \ref{fig2}-\ref{fig4} that
the deviation from SM of the anomalous cross sections including
$\frac{k_{0}^{W}}{\Lambda^{2}}$ is larger than those of containing
$\frac{k_{c}^{W}}{\Lambda^{2}}$ and $\frac{k_{2}^{m}}{\Lambda^{2}}$.
Hence, sensitivities on the coupling $\frac{k_{0}^{W}}{\Lambda^{2}}$
are expected to be more restrictive than the sensitivities on
$\frac{k_{c}^{W}}{\Lambda^{2}}$ and $\frac{k_{2}^{m}}{\Lambda^{2}}$.

\begin{figure}
\includegraphics [width=0.8\columnwidth]{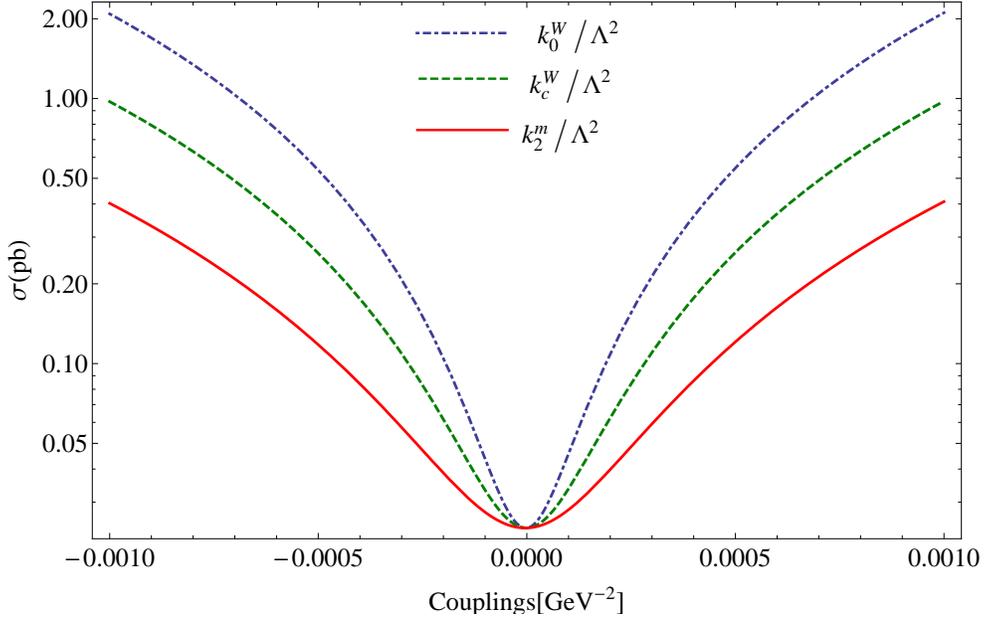}
\caption{The total cross sections as function of anomalous
$\frac{k_{0}^{W}}{\Lambda^{2}}$, $\frac{k_{c}^{W}}{\Lambda^{2}}$ and
$\frac{k_{2}^{m}}{\Lambda^{2}}$ couplings for the  process $\gamma
\gamma\rightarrow W^{+} W^{-}Z $ at the CLIC with $\sqrt{s}=0.5$
TeV. \label{fig6}}
\end{figure}

\begin{figure}
\includegraphics [width=0.8\columnwidth]{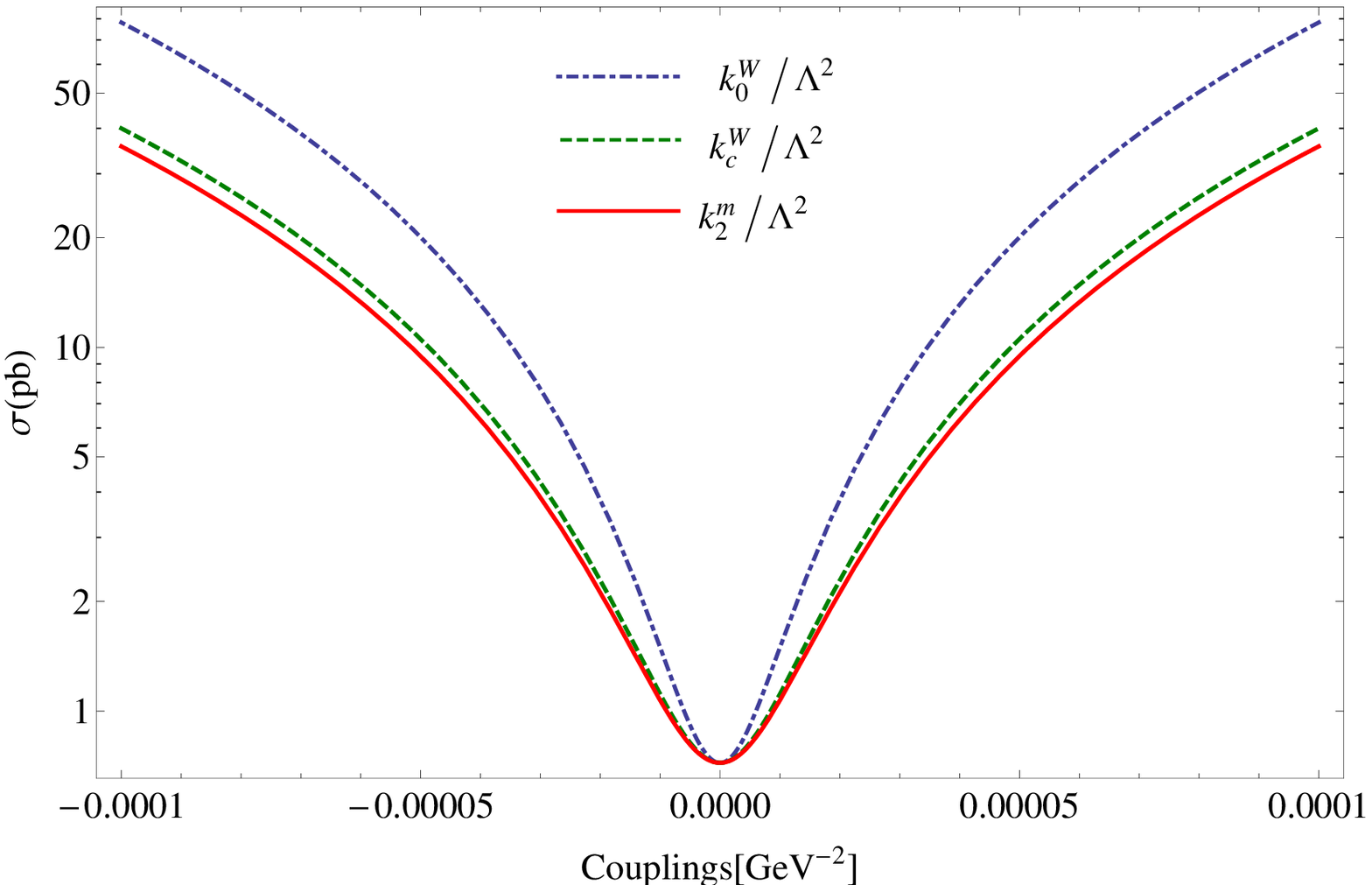}
\caption{The same as Fig. 6 but for $\sqrt{s}=1.5$ TeV.
\label{fig7}}
\end{figure}

\begin{figure}
\includegraphics [width=0.8\columnwidth]{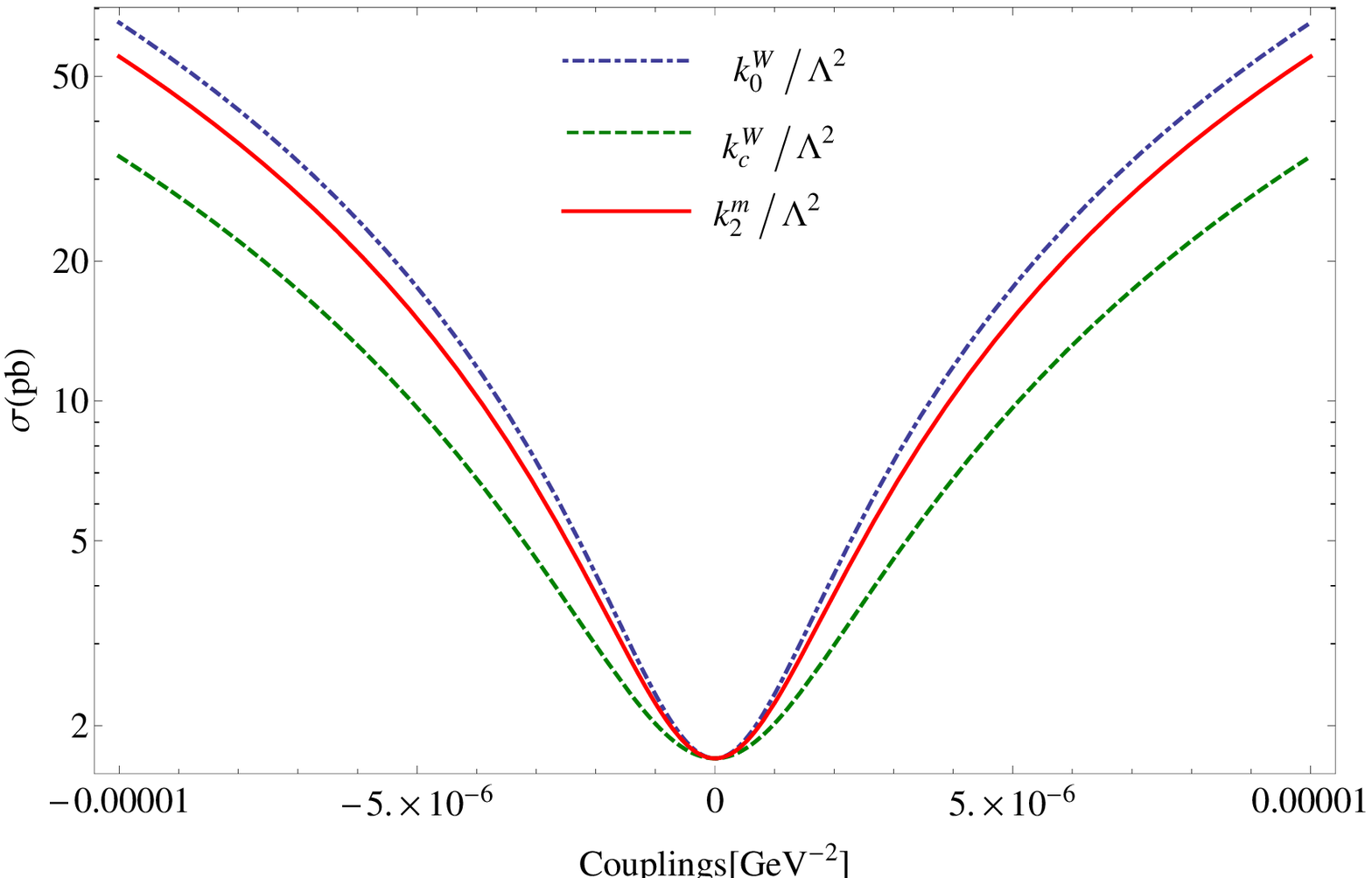}
\caption{The same as Fig. 6 but for $\sqrt{s}=3$ TeV.
\label{fig8}}
\end{figure}

\begin{figure}
\includegraphics [width=0.8\columnwidth] {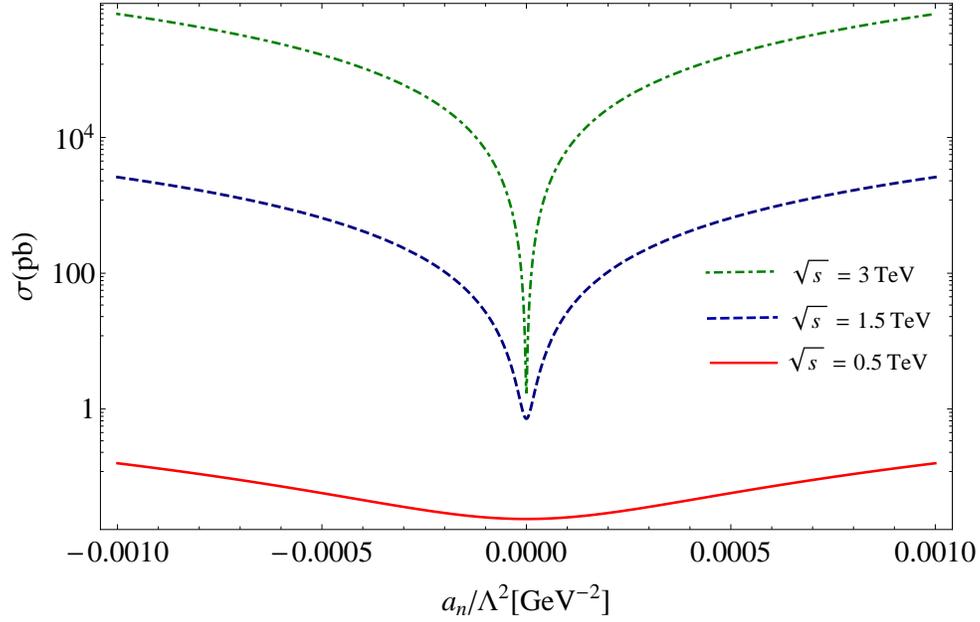}
\caption{The total cross sections as function of anomalous
$\frac{a_{n}}{\Lambda^{2}}$ coupling for the process $\gamma
\gamma\rightarrow W^{+} W^{-}Z $ at the CLIC with $\sqrt{s}=0.5,
1.5$ and $3$ TeV. \label{fig9}}
\end{figure}

The total cross section for the $\gamma \gamma\rightarrow W^{+}
W^{-}Z$ process has been calculated by using real photon spectrum
produced by Compton backscattering of laser beam off the high energy
electron beam. In Figs. \ref{fig6}-\ref{fig8}, we plot the total
cross section of the process $\gamma \gamma\rightarrow W^{+} W^{-}Z$
as a function of anomalous couplings for $\sqrt{s}=0.5,1.5$ and $3$
TeV energies. The total cross section depending on the anomalous
$\frac{a_{n}}{\Lambda^{2}}$ of the process $\gamma \gamma\rightarrow
W^{+} W^{-}Z$ for the three center of mass energies are plotted in
Fig. \ref{fig9}.

\begin{figure}
\includegraphics  [width=0.8\columnwidth]{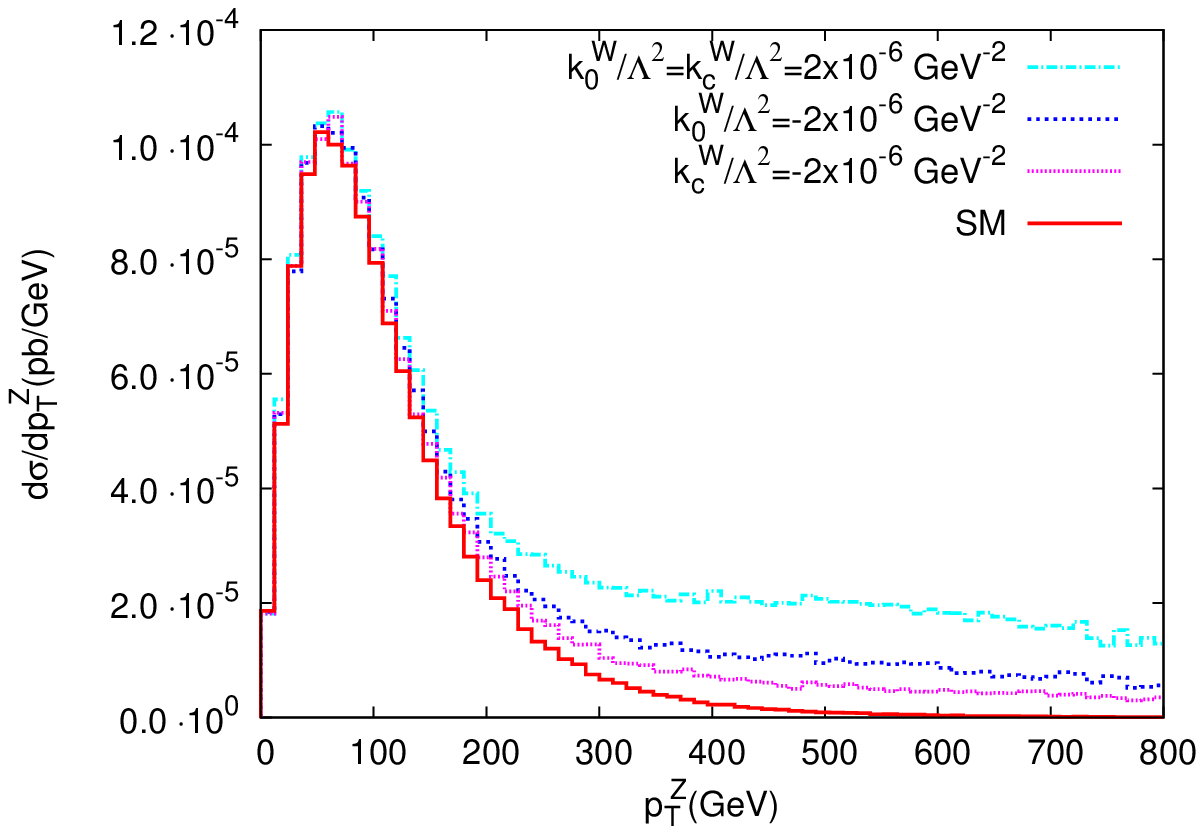}
\caption{The transverse momentum distributions of $Z$ boson in the
final states using anomalous $\frac{k_{0}^{W}}{\Lambda^{2}}$ and
$\frac{k_{c}^{W}}{\Lambda^{2}}$ couplings for the processes
$e^{+}e^{-}\rightarrow e^{+}\gamma^{*} \gamma^{*}e^{-}\rightarrow
e^{+}W^{+} W^{-} Z e^{-}$ at $\sqrt{s}=3$ TeV. \label{fig10}}
\end{figure}

\begin{figure}
\includegraphics [width=0.8\columnwidth]{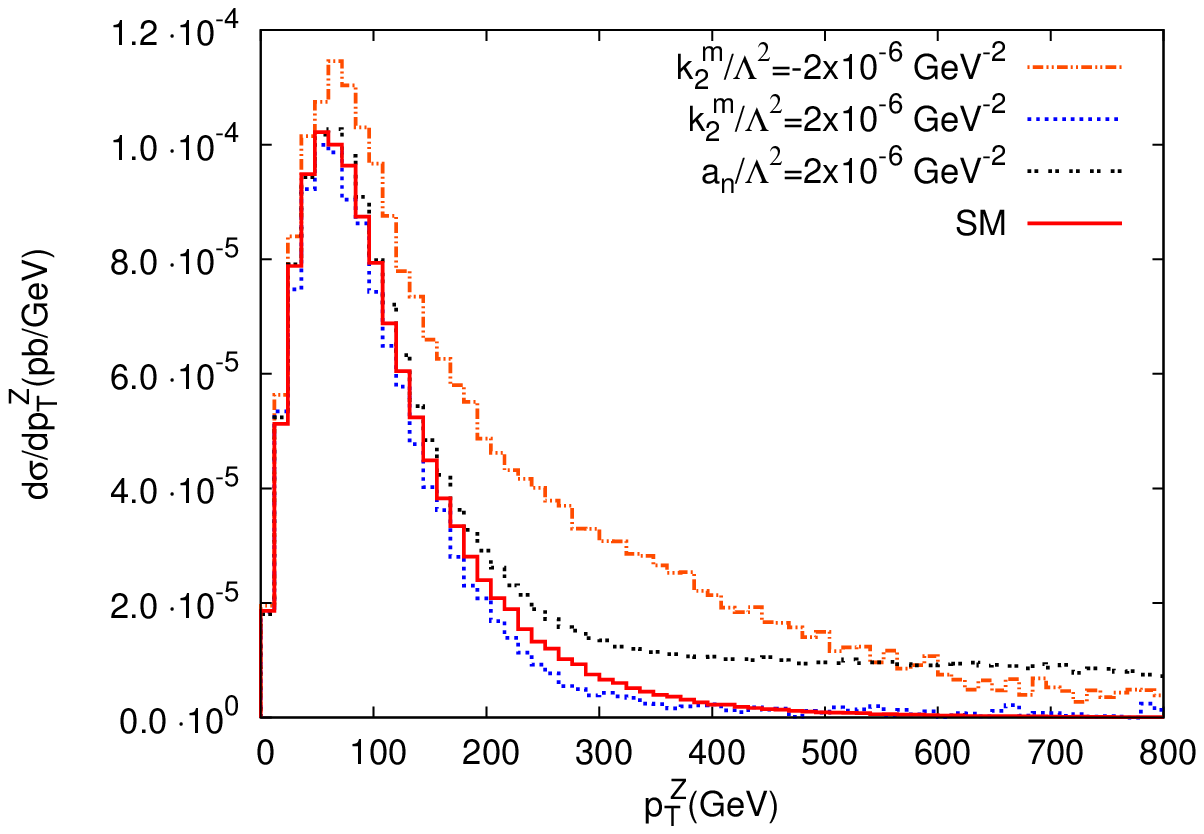}
\caption{The transverse momentum distributions of $Z$ boson in the
final states using anomalous $\frac{k_{2}^{m}}{\Lambda^{2}}$ and
$\frac{a_{n}}{\Lambda^{2}}$ couplings for the processes
$e^{+}e^{-}\rightarrow e^{+}\gamma^{*} \gamma^{*}e^{-}\rightarrow
e^{+}W^{+} W^{-} Z e^{-}$ at $\sqrt{s}=3$ TeV. \label{fig11}}
\end{figure}

\begin{figure}
\includegraphics  [width=0.8\columnwidth]{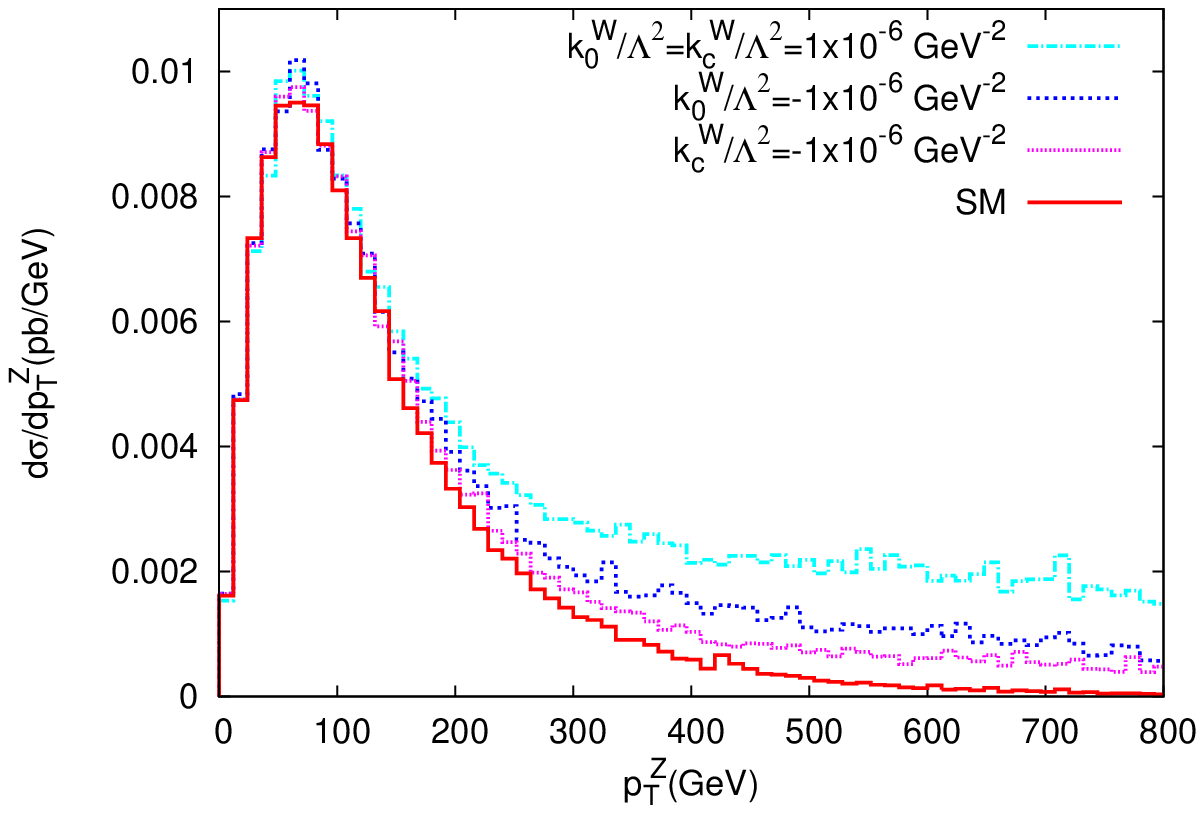}
\caption{The transverse momentum distributions of $Z$ boson in the
final states using anomalous $\frac{k_{0}^{W}}{\Lambda^{2}}$ and
$\frac{k_{c}^{W}}{\Lambda^{2}}$ couplings for the processes $\gamma
\gamma\rightarrow W^{+} W^{-}Z$ at $\sqrt{s}=3$ TeV. \label{fig12}}
\end{figure}

\begin{figure}
\includegraphics [width=0.8\columnwidth]{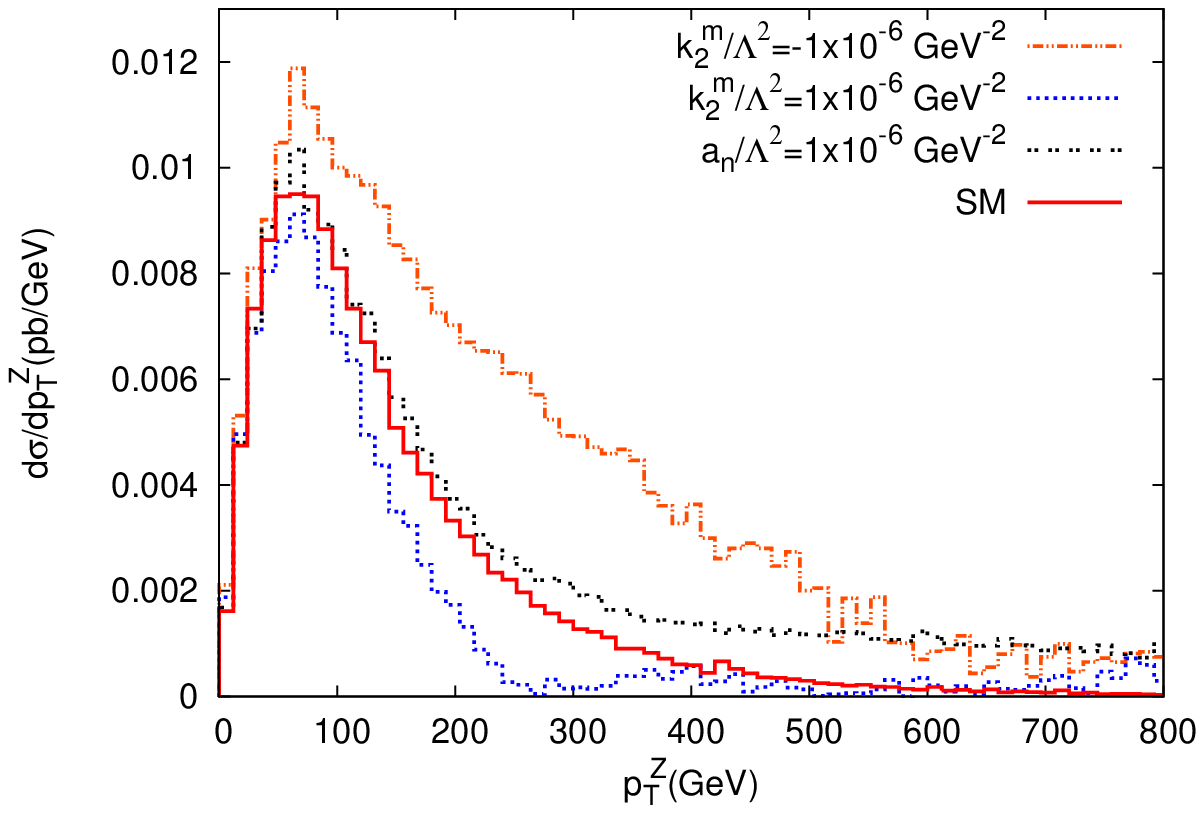}
\caption{The transverse momentum distributions of $Z$ boson in the
final states using anomalous $\frac{k_{2}^{m}}{\Lambda^{2}}$ and
$\frac{a_{n}}{\Lambda^{2}}$ couplings for the processes $\gamma
\gamma\rightarrow W^{+} W^{-}Z$ at $\sqrt{s}=3$ TeV. \label{fig13}}
\end{figure}

\begin{figure}
\includegraphics  [width=0.8\columnwidth]{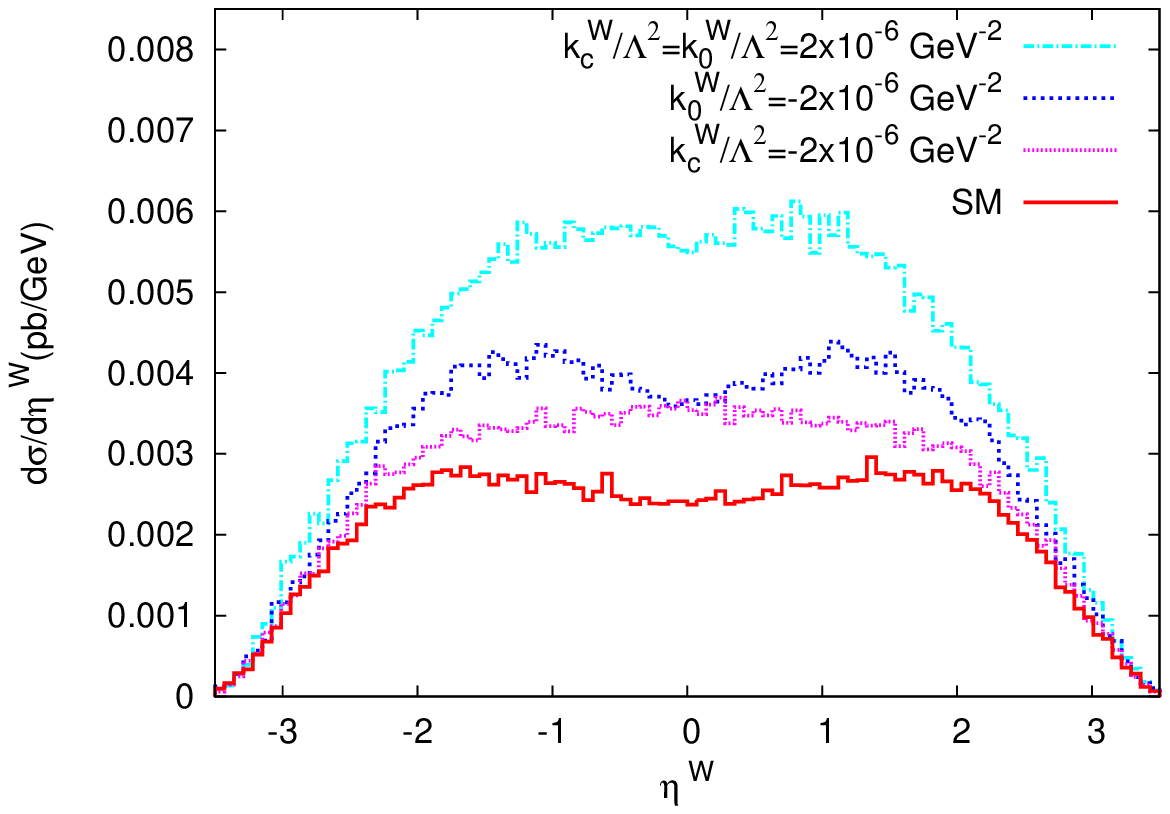}
\caption{The rapidity distributions of $W^{+}$ boson in the final
states using anomalous $\frac{k_{0}^{W}}{\Lambda^{2}}$ and
$\frac{k_{c}^{W}}{\Lambda^{2}}$ couplings for the processes
$e^{+}e^{-}\rightarrow e^{+}\gamma^{*} \gamma^{*}e^{-}\rightarrow
e^{+}W^{+} W^{-} Z e^{-}$ at $\sqrt{s}=3$ TeV. \label{fig14}}
\end{figure}

\begin{figure}
\includegraphics  [width=0.8\columnwidth]{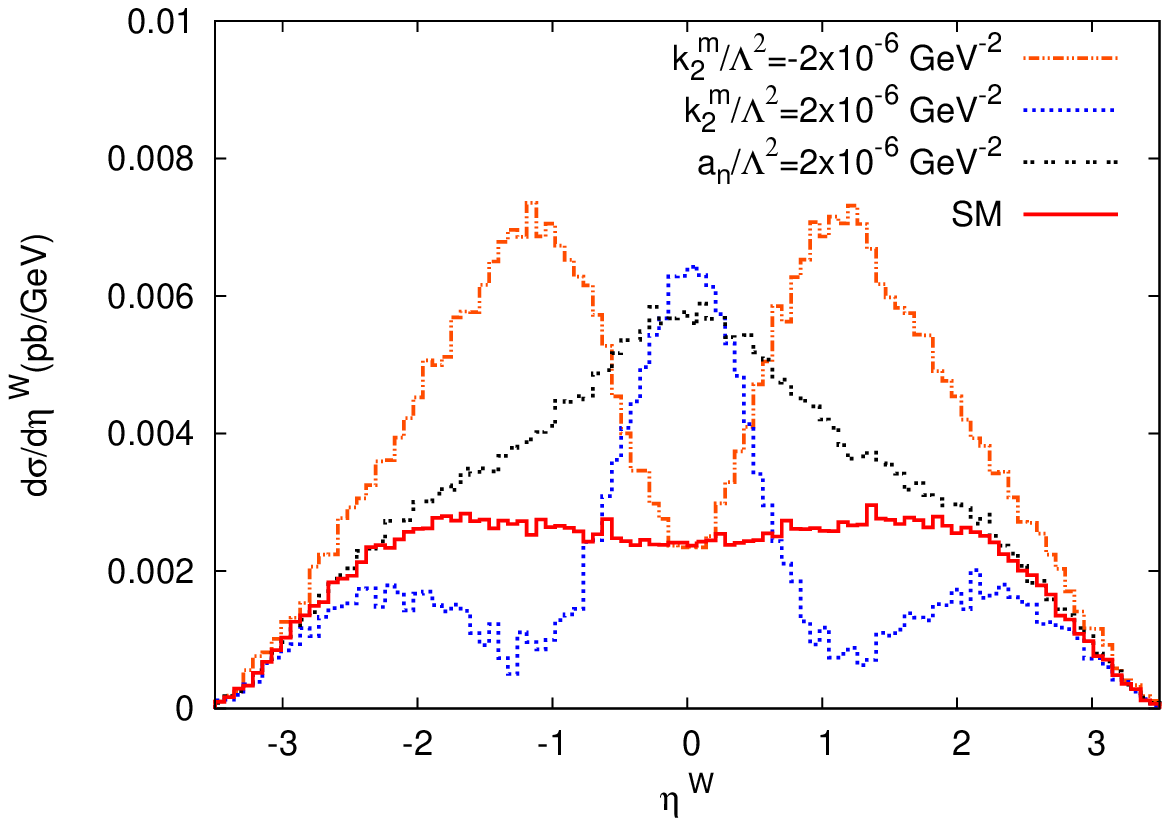}
\caption{The rapidity distributions of $W^{+}$ boson in the final
states using anomalous $\frac{k_{2}^{m}}{\Lambda^{2}}$ and
$\frac{a_{n}}{\Lambda^{2}}$ couplings for the processes
$e^{+}e^{-}\rightarrow e^{+}\gamma^{*} \gamma^{*}e^{-}\rightarrow
e^{+}W^{+} W^{-} Z e^{-}$ at $\sqrt{s}=3$ TeV. \label{fig15}}
\end{figure}

\begin{figure}
\includegraphics  [width=0.8\columnwidth]{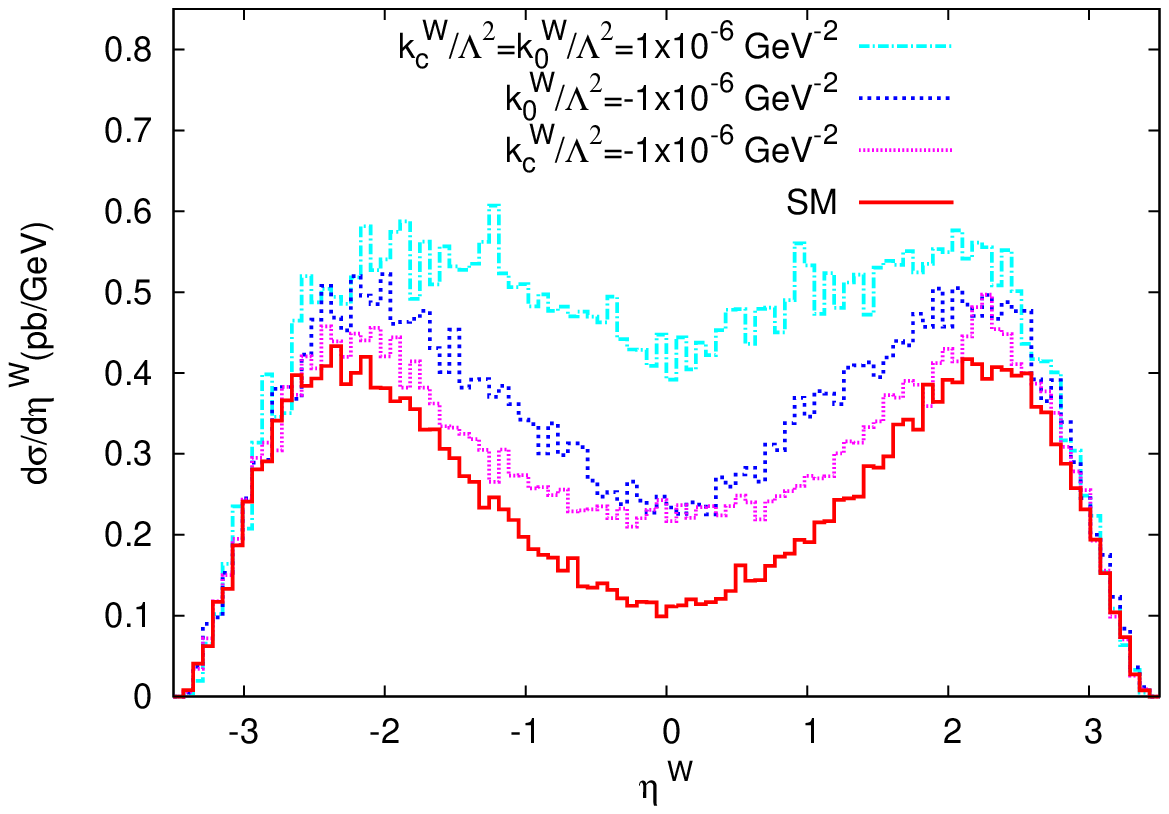}
\caption{The rapidity distributions of $W^{+}$ boson in the final
states using anomalous $\frac{k_{0}^{W}}{\Lambda^{2}}$ and
$\frac{k_{c}^{W}}{\Lambda^{2}}$ couplings for the processes $\gamma
\gamma\rightarrow W^{+} W^{-}Z$ at $\sqrt{s}=3$ TeV. \label{fig16}}
\end{figure}

\begin{figure}
\includegraphics  [width=0.8\columnwidth]{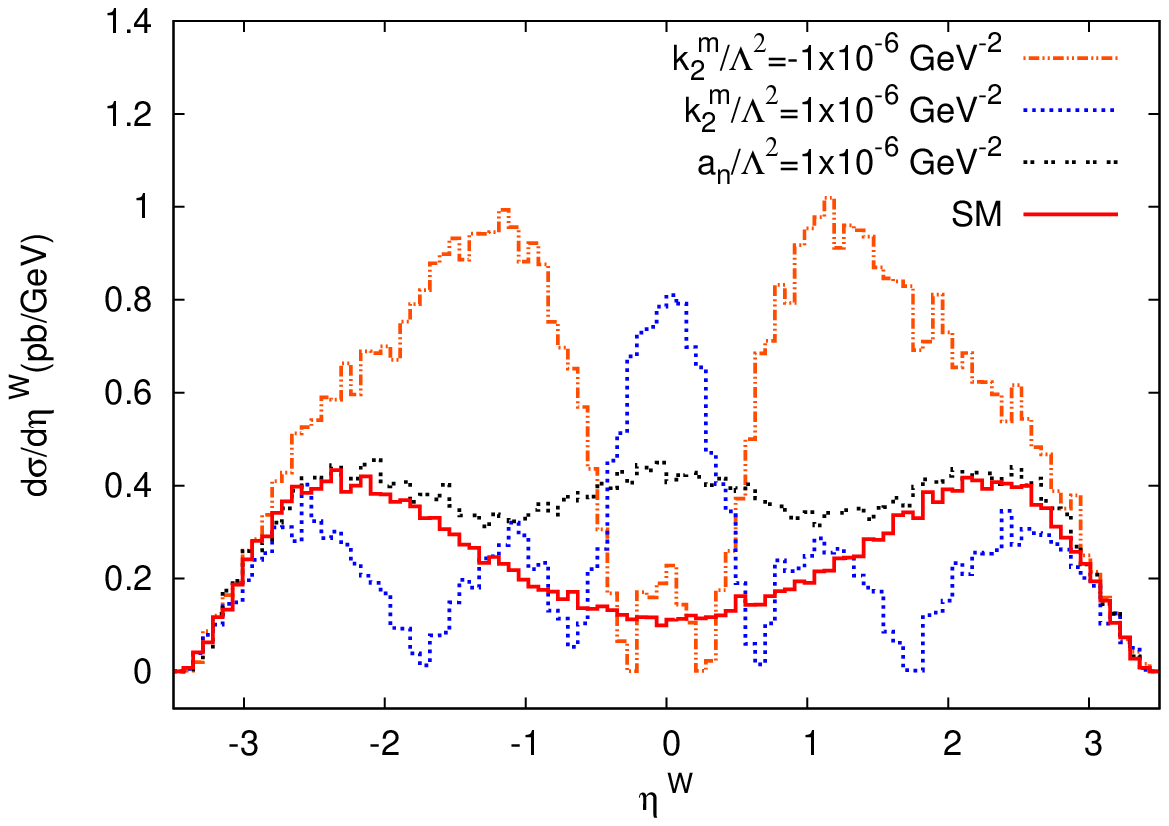}
\caption{The rapidity distributions of $W^{+}$ boson in the final
states using anomalous $\frac{k_{2}^{m}}{\Lambda^{2}}$ and
$\frac{a_{n}}{\Lambda^{2}}$ couplings for the processes $\gamma
\gamma\rightarrow W^{+} W^{-}Z$ at $\sqrt{s}=3$ TeV. \label{fig17}}
\end{figure}

\begin{figure}
\includegraphics  [width=0.8\columnwidth]{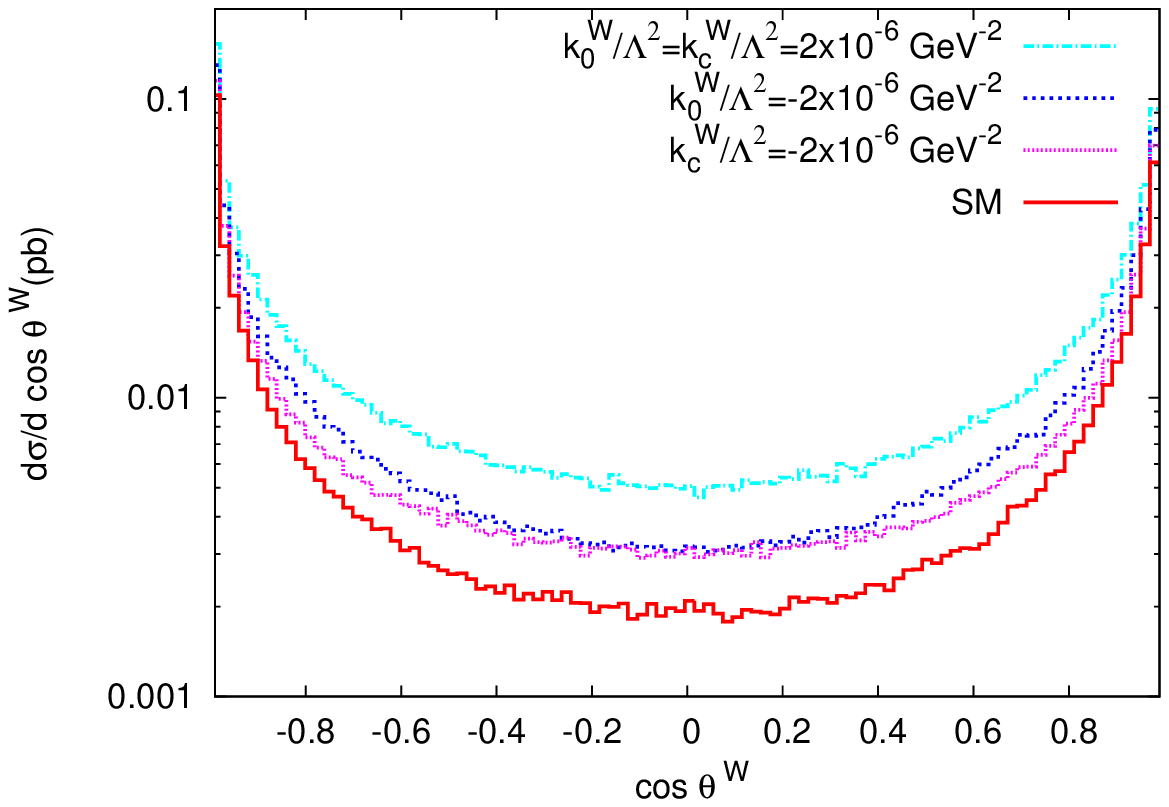}
\caption{The angular distributions of $W^{+}$ boson in the final
states using anomalous $\frac{k_{0}^{W}}{\Lambda^{2}}$ and
$\frac{k_{c}^{W}}{\Lambda^{2}}$ couplings for the processes
$e^{+}e^{-}\rightarrow e^{+}\gamma^{*} \gamma^{*}e^{-}\rightarrow
e^{+}W^{+} W^{-} Z e^{-}$ at $\sqrt{s}=3$ TeV. \label{fig18}}
\end{figure}

\begin{figure}
\includegraphics  [width=0.8\columnwidth]{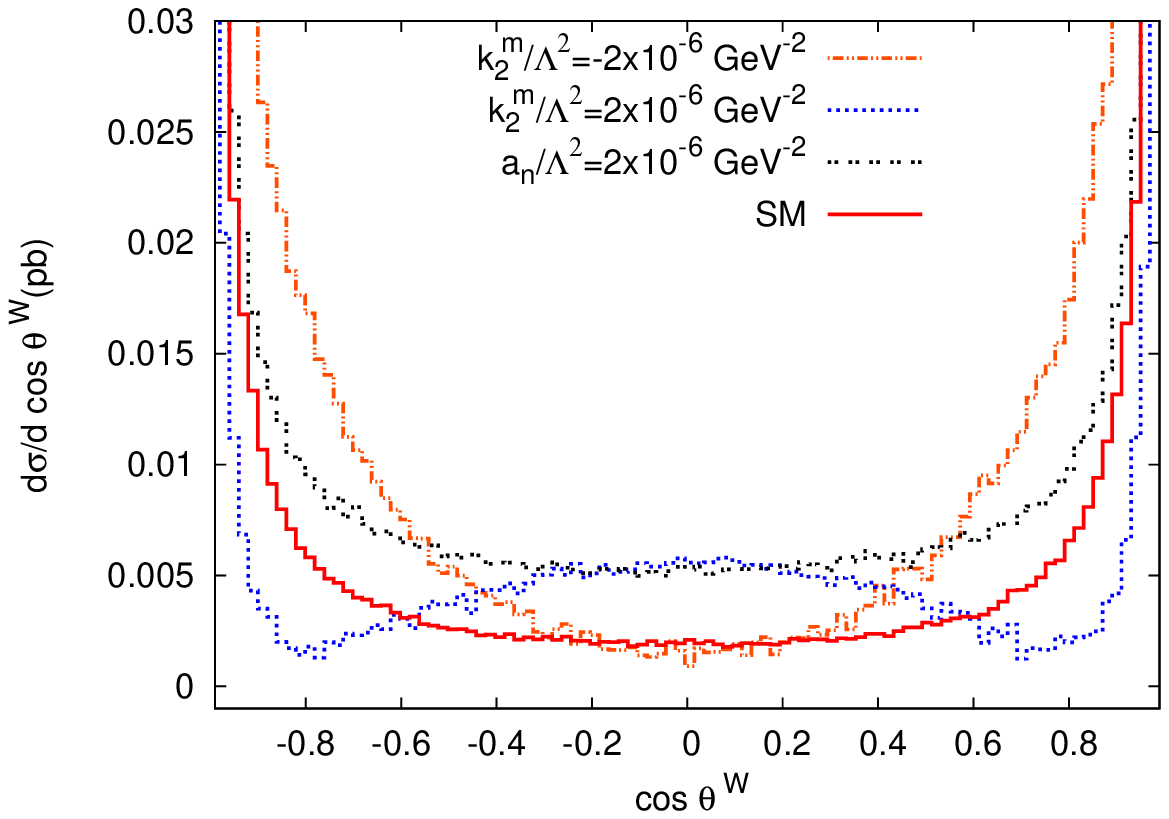}
\caption{The angular distributions of $W^{+}$ boson in the final
states using anomalous $\frac{k_{2}^{m}}{\Lambda^{2}}$ and
$\frac{a_{n}}{\Lambda^{2}}$ couplings for the processes
$e^{+}e^{-}\rightarrow e^{+}\gamma^{*} \gamma^{*}e^{-}\rightarrow
e^{+}W^{+} W^{-} Z e^{-}$ at $\sqrt{s}=3$ TeV. \label{fig19}}
\end{figure}

\begin{figure}
\includegraphics  [width=0.8\columnwidth]{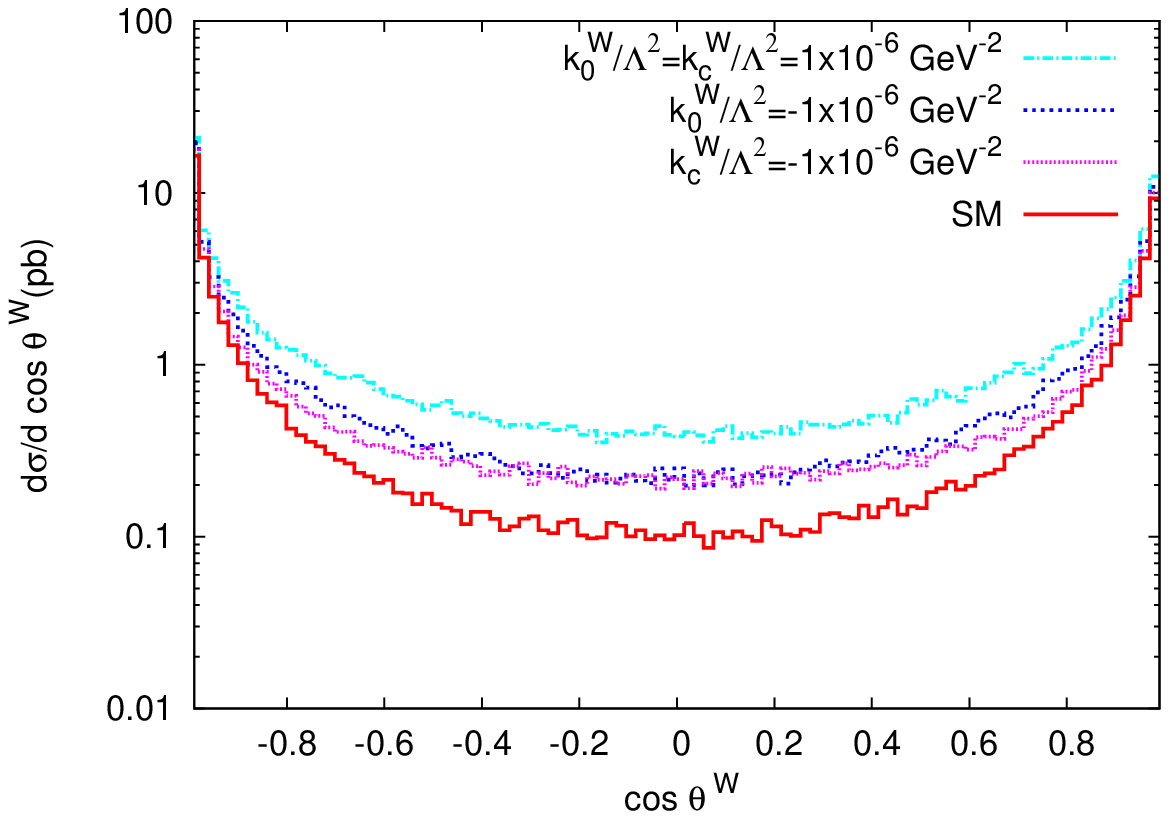}
\caption{The angular distributions of $W^{+}$ boson in the final
states using anomalous $\frac{k_{0}^{W}}{\Lambda^{2}}$ and
$\frac{k_{c}^{W}}{\Lambda^{2}}$ couplings for the processes $\gamma
\gamma\rightarrow W^{+} W^{-}Z$ at $\sqrt{s}=3$ TeV. \label{fig20}}
\end{figure}

\begin{figure}
\includegraphics  [width=0.8\columnwidth]{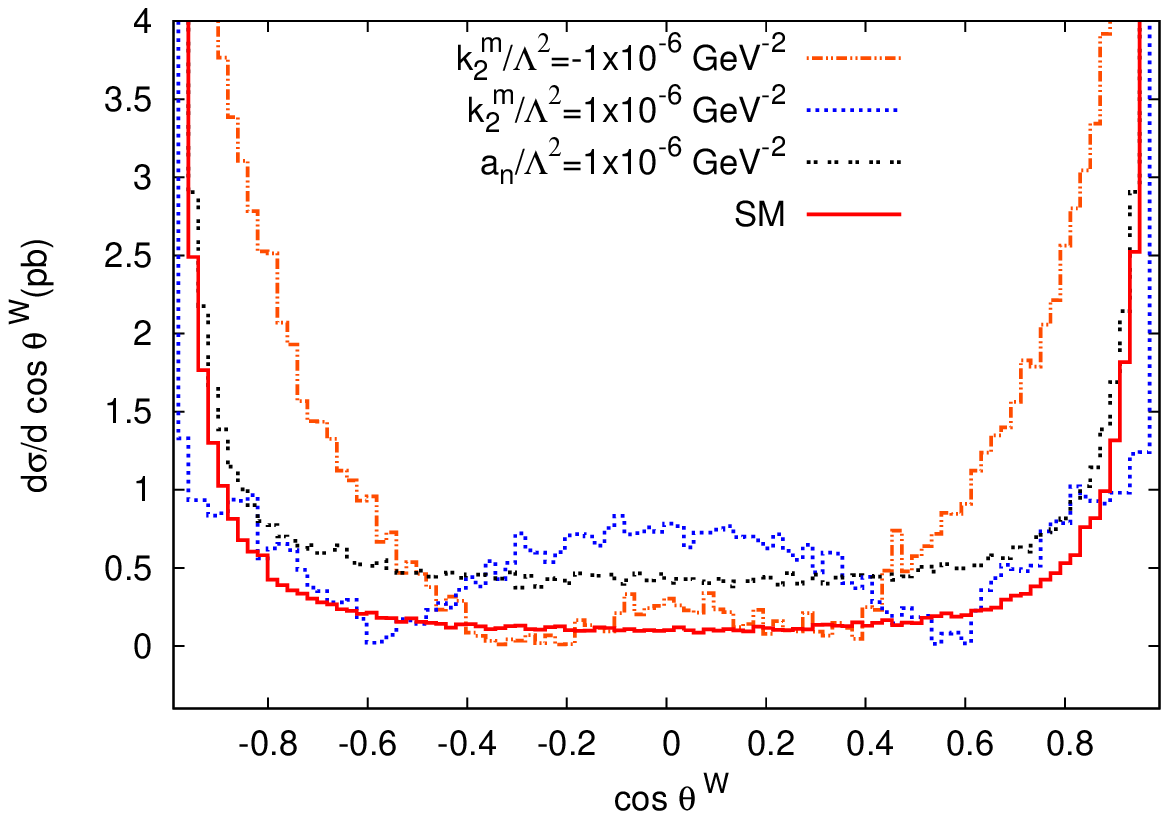}
\caption{The angular distributions of $W^{+}$ boson in the final
states using anomalous $\frac{k_{2}^{m}}{\Lambda^{2}}$ and
$\frac{a_{n}}{\Lambda^{2}}$ couplings for the processes $\gamma
\gamma\rightarrow W^{+} W^{-}Z$ at $\sqrt{s}=3$ TeV. \label{fig21}}
\end{figure}

The kinematical distributions of final state particles can give
further information about how we can separate among the different
anomalous interactions. In this context, some distributions of the
final state $W$ and $Z$ bosons are plotted for illustrative purposes
using close to sensitivity of the anomalous couplings
$\frac{k_{0}^{W}}{\Lambda^{2}}$, $\frac{k_{c}^{W}}{\Lambda^{2}}$,
$\frac{k_{2}^{m}}{\Lambda^{2}}$ and $\frac{a_{n}}{\Lambda^{2}}$ in
Figs. \ref{fig10}-\ref{fig21}. We show the transverse momentum
distributions of $Z$ boson in the final states using
$\frac{k_{0}^{W}}{\Lambda^{2}}$, $\frac{k_{c}^{W}}{\Lambda^{2}}$
anomalous couplings in Fig. \ref{fig10} and using
$\frac{k_{2}^{m}}{\Lambda^{2}}$ and $\frac{a_{n}}{\Lambda^{2}}$
anomalous couplings in Fig. \ref{fig11} for the processes
$e^{+}e^{-}\rightarrow e^{+}\gamma^{*} \gamma^{*} e^{-} \rightarrow
e^{+} W^{+} W^{-} Z e^{-}$ at $\sqrt{s}=3$ TeV. Similarly, the
transverse momentum distributions for the $Z$ boson in the final
states of the process $\gamma \gamma\rightarrow W^{+} W^{-}Z$ are
given in Figs. \ref{fig12}-\ref{fig13}. From this figures, we can
separately observe the deviation of new physics induced by nonzero
anomalous quartic CP-conserving and CP-violating couplings apart
from SM background which is apparent at high $p_T$ region of the $Z$
bosons in the final states. The momentum dependence of the anomalous
cross sections including the $WW Z\gamma$ vertices is higher than
that of SM background cross section which causes the apparent
deviation at high $p_T$ region.

We plot the rapidity distributions of the $W^{+}$ boson for two
processes using the anomalous couplings
$\frac{k_{0}^{W}}{\Lambda^{2}}$, $\frac{k_{c}^{W}}{\Lambda^{2}}$,
$\frac{k_{2}^{m}}{\Lambda^{2}}$ and $\frac{a_{n}}{\Lambda^{2}}$ for
$\sqrt{s}=3$ TeV in Figs. \ref{fig14}-\ref{fig17}. Figs.
\ref{fig14}-\ref{fig17} show that the rapidity distributions of the
final state $W^{+}$ boson from the new physics signals and SM
background are located generally in the range of $|\eta^{W}|<2.5$.
Furthermore, we can easily discern the difference between positive
and negative values of the coupling $\frac{k_{2}^{m}}{\Lambda^{2}}$.
Especially, as can be seen from Fig. $15$, the anomalous
interactions for $\frac{a_{n}}{\Lambda^{2}}$ coupling cause the
production of more $W^{+}$ bosons in the central region.


In order to distinguish the different anomalous couplings with the
SM, we illustrate the $cos\theta^W$ distributions of $W^{+}$ for two
processes where $\theta^W$ is polar angle of $W^{+}$ with respect to
the beam pipe. We show the $cos\theta^W$ distributions with the
anomalous couplings $\frac{k_{0}^{W}}{\Lambda^{2}}$,
$\frac{k_{c}^{W}}{\Lambda^{2}}$ and SM background in Fig.
\ref{fig18}, using $\frac{k_{2}^{m}}{\Lambda^{2}}$ and
$\frac{a_{n}}{\Lambda^{2}}$ couplings in Fig. \ref{fig19} for
$e^{+}e^{-}\rightarrow e^{+}\gamma^{*} \gamma^{*} e^{-} \rightarrow
e^{+} W^{+} W^{-} Z e^{-}$ process at $\sqrt{s}=3$ TeV. Similarly,
the $cos\theta^W$ distributions for the $\gamma \gamma\rightarrow
W^{+} W^{-}Z$ process are given in Figs. \ref{fig20}-\ref{fig21}. We
can observe from these distributions that the contributions of
negative and positive values of $\frac{k_{2}^{m}}{\Lambda^{2}}$ can
easily be distinguished in Figs. \ref{fig19} and \ref{fig21}.

In order to probe the sensitivity to the anomalous quartic
$WWZ\gamma$ couplings, we use one and two-dimensional $\chi^{2}$
analysis:
\begin{eqnarray}
\chi^{2}=\left(\frac{\sigma_{SM}-\sigma_{AN}(\frac{k_{0,c}^{W}}{\Lambda^{2}},
\frac{k_{2}^{m}}{\Lambda^{2}},
\frac{a_{n}}{\Lambda^{2}})}{\sigma_{SM}\delta_{stat}}\right)^{2}
\end{eqnarray}
where $\sigma_{AN}$ is the cross section including new physics
effects, $\delta_{stat}=\frac{1}{\sqrt{N}}$ and $N$ is the number of
SM events. The number of events for the processes $\gamma
\gamma\rightarrow W^{+} W^{-}Z $  and $e^{+}e^{-} \rightarrow
e^{+}\gamma^{*} \gamma^{*} e^{-} \rightarrow e^{+} W^{+} W^{-} Z
e^{-}$ are obtained by $N=L_{int} \times \sigma_{SM} \times
BR(Z\rightarrow\ell \bar{\ell})\times BR^{2}(W\rightarrow q
\bar{q}')$ where $L_{int}$ is the integrated luminosity. In
addition, we assume that the the leptonic decay channel of $Z$ boson
with branching ratio is $BR(Z\rightarrow \ell \bar{\ell})=0.067$ and
the hadronic decay channel of $W$ boson with the branching ratio is
$BR(W\rightarrow q \bar{q}')=0.676$. In our calculations, one of the
anomalous quartic couplings is assumed to deviate from their SM
values (the others fixed to zero) at the one-dimensional $\chi^{2}$
analysis, while two anomalous quartic couplings
($\frac{k_{0,c}^{W}}{\Lambda^{2}}$) are assumed to deviate from
their SM values at the two-dimensional $\chi^{2}$ analysis. In this
case, we take into account $\chi^2$ value corresponding to the
number of observable.

In Tables \ref{tab1}-\ref{tab4}, we show $95\%$ C. L. sensitivities
on the anomalous quartic couplings parameters
$\frac{k_{0}^{W}}{\Lambda^{2}}$, $\frac{k_{c}^{W}}{\Lambda^{2}}$ ,
and $\frac{k_{2}^{m}}{\Lambda^{2}}$, $\frac{a_{n}}{\Lambda^{2}}$ for
both two processes at $\sqrt{s}=0.5, 1.5$ and $3$ TeV energies. As
can be seen in Table \ref{tab1}, the process $\gamma
\gamma\rightarrow W^{+} W^{-}Z $ improves the sensitivities of
$\frac{k_{0}^{W}}{\Lambda^{2}}$ and $\frac{k_{c}^{W}}{\Lambda^{2}}$
by up to a factor of $10^{2}$ compared to the LHC \cite{sınır}.
The expected best sensitivities on $\frac{a_{n}}{\Lambda^{2}}$ in
Table \ref{tab2} are far beyond the sensitivities of the existing
LEP. However, we compare our results with the sensitivities of Ref.
\cite{mur}, in which the best sensitivities on
$\frac{k_{0}^{W}}{\Lambda^{2}}$, $\frac{k_{c}^{W}}{\Lambda^{2}}$,
$\frac{k_{2}^{m}}{\Lambda^{2}}$ and $\frac{a_{n}}{\Lambda^{2}}$
couplings by examining the two processes $e^{+}e^{-}\rightarrow
W^{-} W^{+}\gamma$ and $e^{+}e^{-} \rightarrow e^{+}\gamma^{*} e^{-}
\rightarrow e^{+} W^{-} Z \nu_{e}$ at the 3 TeV CLIC are obtained.
We observed that the sensitivities obtained on
$\frac{k_{0}^{W}}{\Lambda^{2}}$ and $\frac{k_{c}^{W}}{\Lambda^{2}}$
are at the same order with those reported in the Ref. \cite{mur}
while sensitivities on $\frac{k_{2}^{m}}{\Lambda^{2}}$ and
$\frac{a_{n}}{\Lambda^{2}}$ are 2 and 5 times better than the
sensitivities calculated in Ref. \cite{mur}, respectively. Our
sensitivities on $\frac{k_{2}^{m}}{\Lambda^{2}}$ can set more
stringent sensitive by two orders of magnitude with respect to the
best sensitivity derived from $WZ\gamma$ production at the LHC with
$\sqrt{s}=14$ TeV and the integrated luminosity of $L=200$ fb$^{-1}$
\cite{lhc1}.

The $\gamma^{*} \gamma^{*}$ collision of CLIC with $\sqrt{s}=3$ TeV
and $L_{int} = 590$ fb$^{-1}$ investigates the CP-conserving and
CP-violating anomalous $WWZ\gamma$ coupling with a far better than
the experiments sensitivities. One can see from Table \ref{tab3}
that the sensitivities on the anomalous couplings
$\frac{k_{0}^{W}}{\Lambda^{2}}$ and $\frac{k_{c}^{W}}{\Lambda^{2}}$
are calculated as $[-1.09;\, 1.09]\times 10^{-6}$ GeV$^{-2}$ and
$[-1.54;\, 1.54]\times 10^{-6}$ GeV$^{-2}$ which are an order of
magnitude better than both $\frac{k_{0}^{W}}{\Lambda^{2}}$ and
$\frac{k_{c}^{W}}{\Lambda^{2}}$ couplings. As shown in Table
\ref{tab4}, the best sensitivities on $\frac{a_{n}}{\Lambda^{2}}$
coupling through the process $e^{+}e^{-} \rightarrow e^{+}\gamma^{*}
\gamma^{*} e^{-} \rightarrow e^{+} W^{+} W^{-} Z e^{-}$ are obtained
as $[-1.04;\, 1.04]\times 10^{-6}$ GeV$^{-2}$ which are more
stringent sensitivity by five orders of magnitude with respect to
LEP results. Anomalous $\frac{k_{0}^{W}}{\Lambda^{2}}$ and
$\frac{k_{c}^{W}}{\Lambda^{2}}$ couplings calculated with the help
of the process $e^{+}e^{-}\rightarrow e^{+}\gamma^{*}
\gamma^{*}e^{-}\rightarrow e^{+}W^{+} W^{-} Z e^{-}$ are less
sensitive than the results of Ref. \cite{mur}. On the other hand,
our $\frac{k_{2}^{m}}{\Lambda^{2}}$ and $\frac{a_{n}}{\Lambda^{2}}$
couplings obtained from this process have similar sensitivities as
Ref. \cite{mur}.

\begin{figure}
\includegraphics [width=0.8\columnwidth]{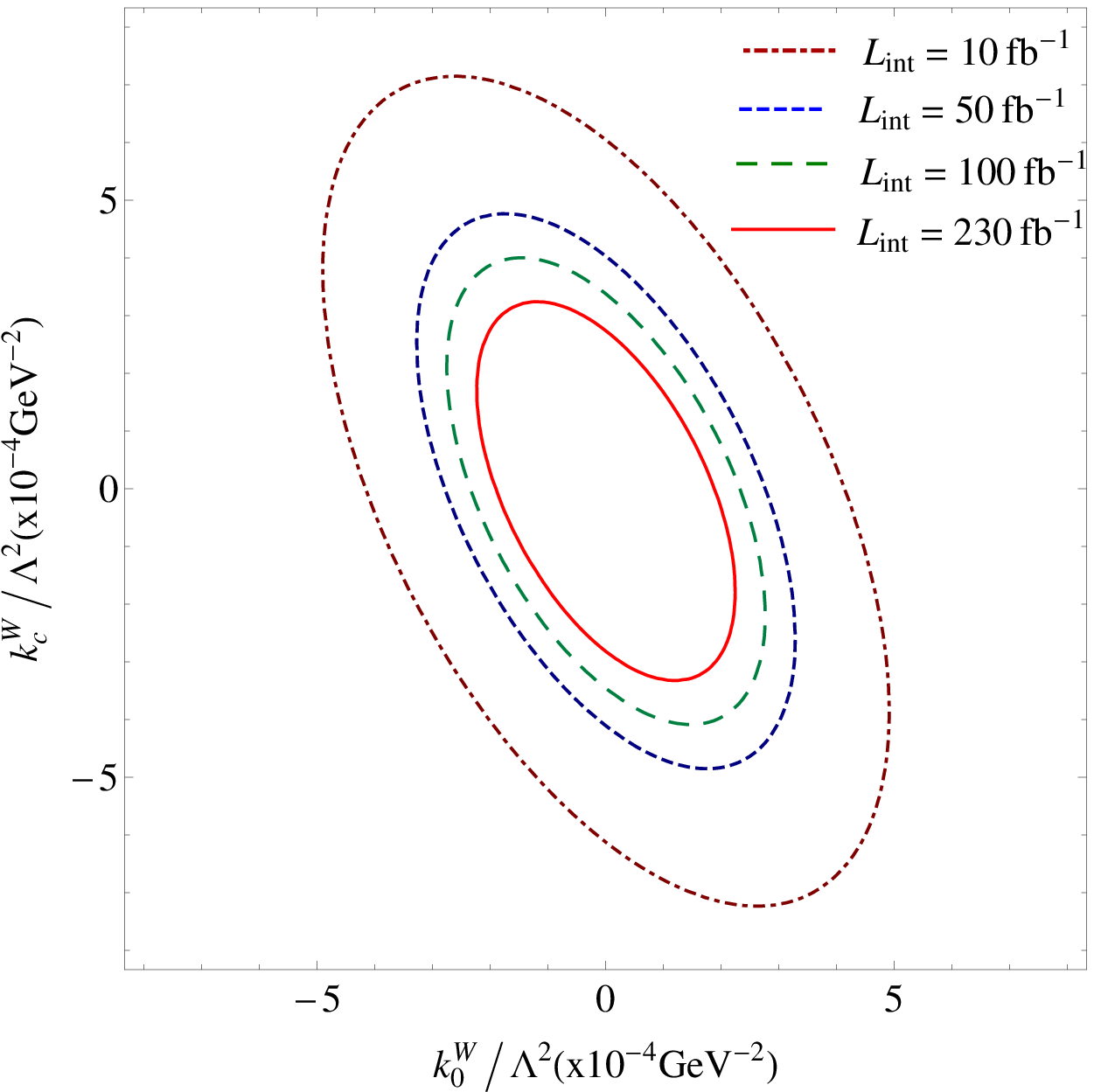}
\caption{$95\%$ C.L. contours for anomalous
$\frac{k_{0}^{W}}{\Lambda^{2}}$ and $\frac{k_{c}^{W}}{\Lambda^{2}}$
couplings for the process $e^{+}e^{-}\rightarrow e^{+}\gamma^{*}
\gamma^{*}e^{-}\rightarrow e^{+}W^{+} W^{-} Z e^{-}$ at the CLIC
with $\sqrt{s}=0.5$ TeV. \label{fig22}}
\end{figure}

\begin{figure}
\includegraphics[width=0.8\columnwidth]{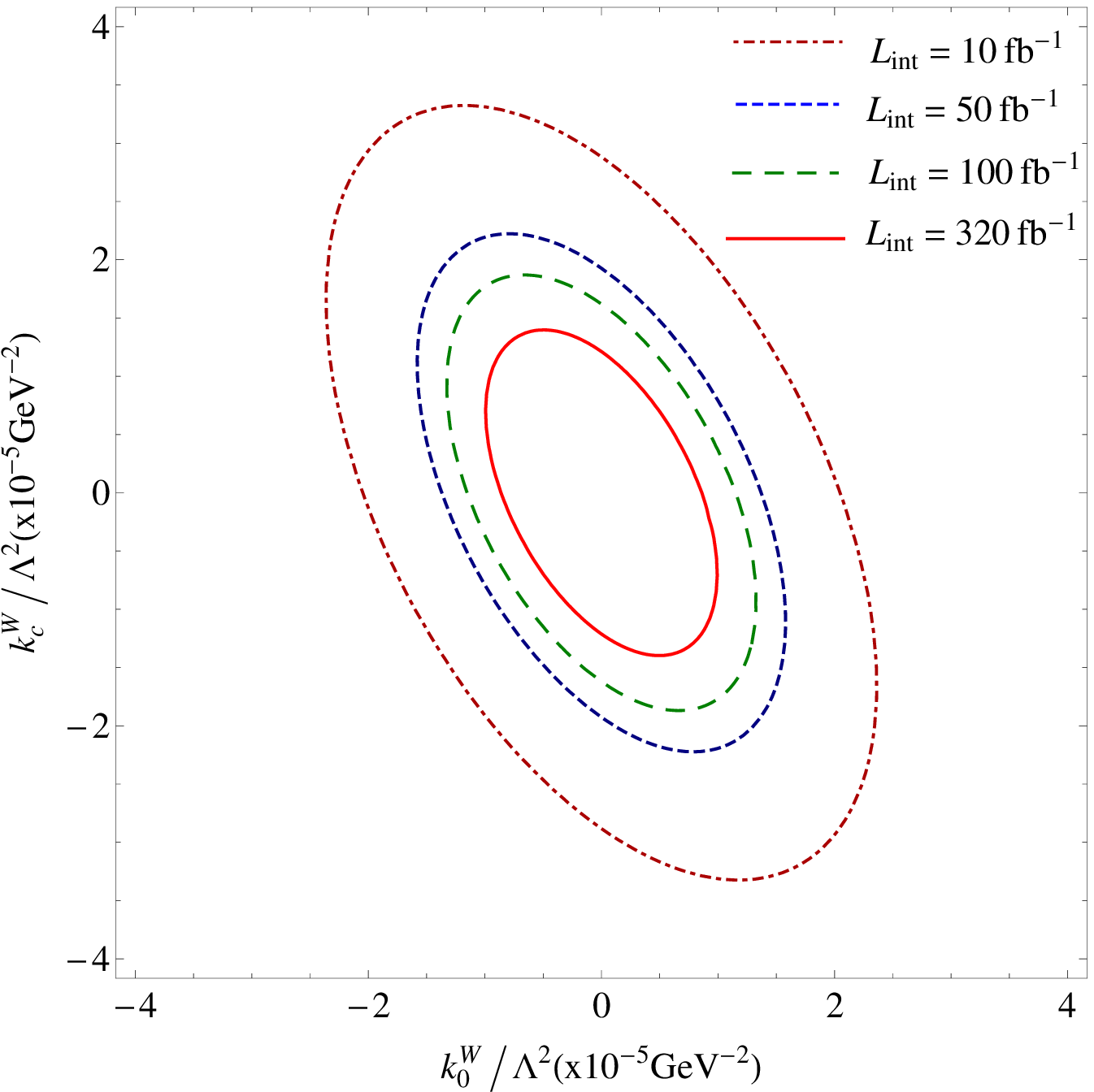}
\caption{The same as Fig. 22 but for $\sqrt{s}=1.5$ TeV.
\label{fig23}}
\end{figure}

\begin{figure}
\includegraphics [width=0.8\columnwidth]{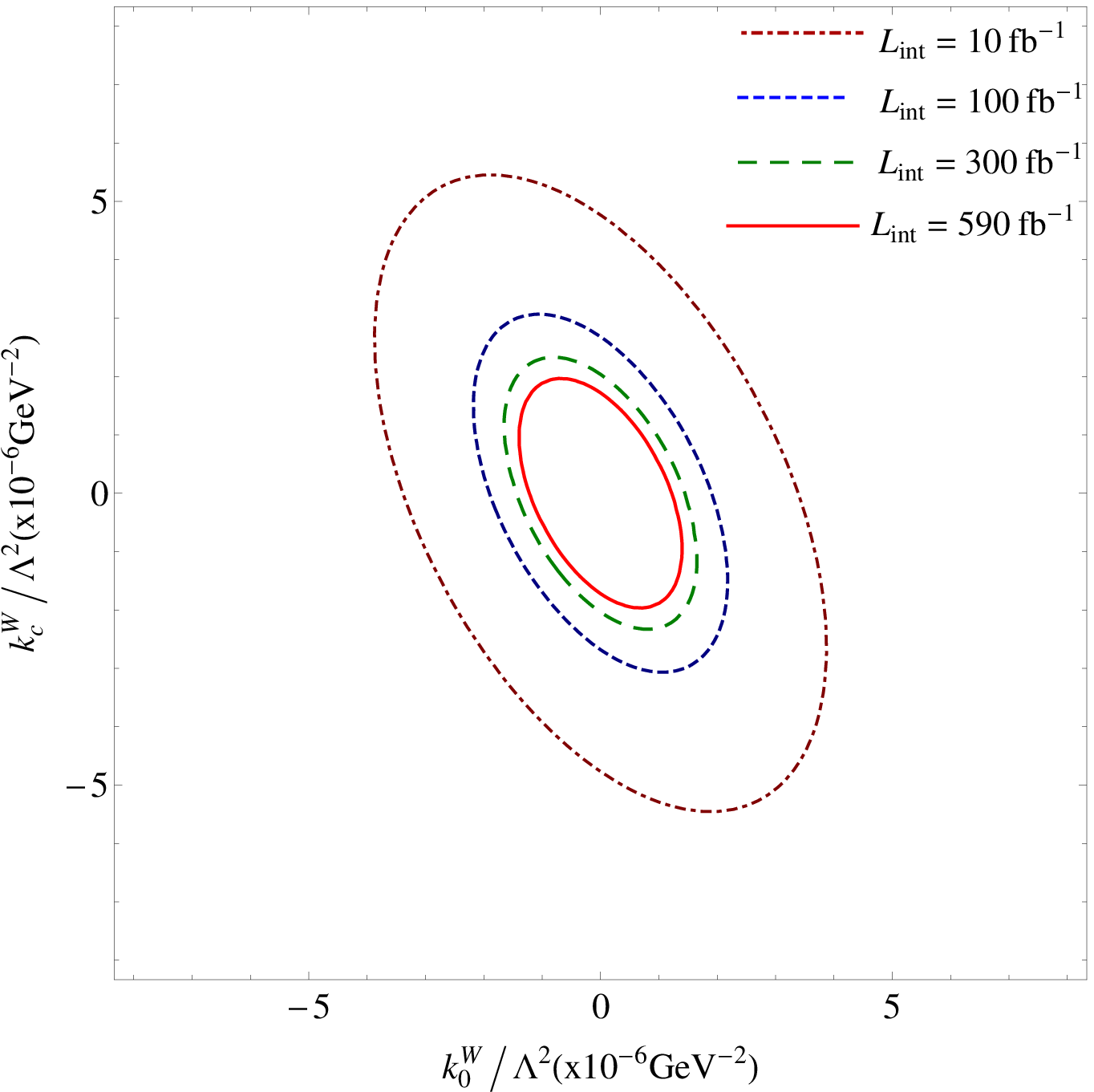}
\caption{The same as Fig. 22 but for $\sqrt{s}=3$ TeV.
\label{fig24}}
\end{figure}

\begin{figure}
\includegraphics [width=0.8\columnwidth]{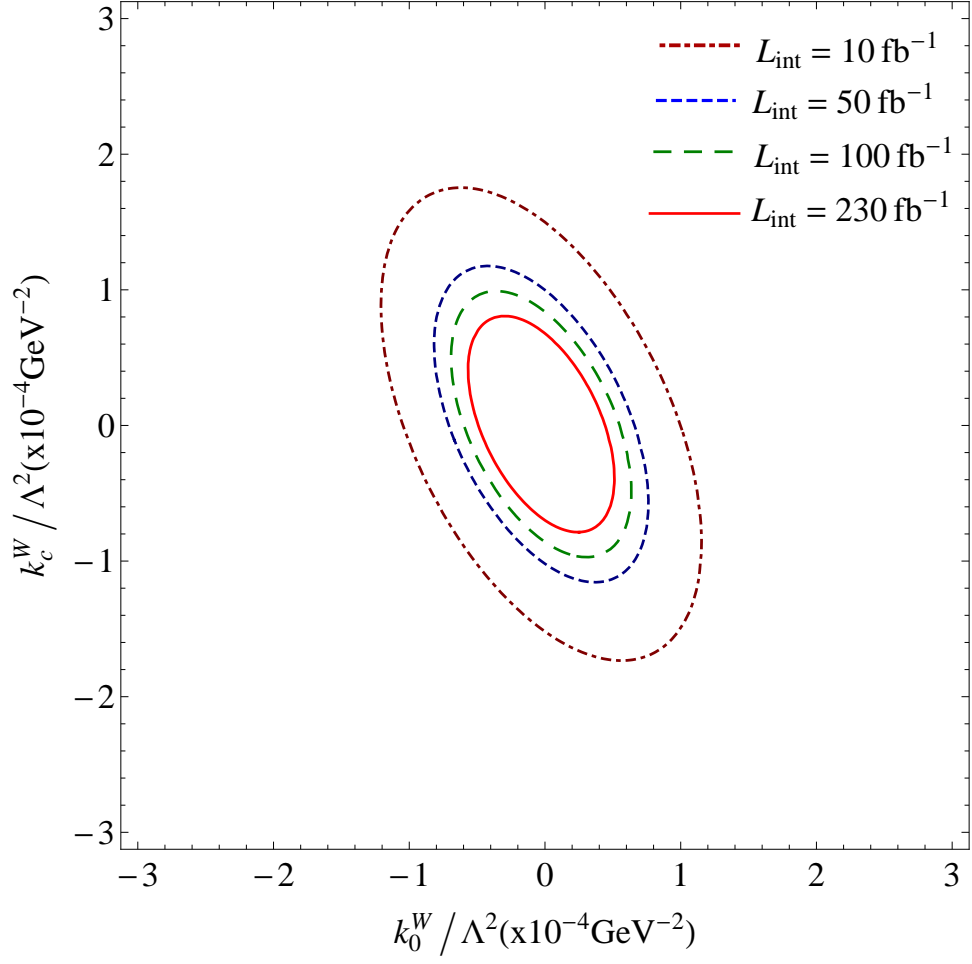}
\caption{$95\%$ C.L. contours for anomalous
$\frac{k_{0}^{W}}{\Lambda^{2}}$ and $\frac{k_{c}^{W}}{\Lambda^{2}}$
couplings for the process $\gamma \gamma\rightarrow W^{+} W^{-}Z$ at
the CLIC with $\sqrt{s}=0.5$ TeV. \label{fig25}}
\end{figure}

\begin{figure}
\includegraphics[width=0.8\columnwidth]{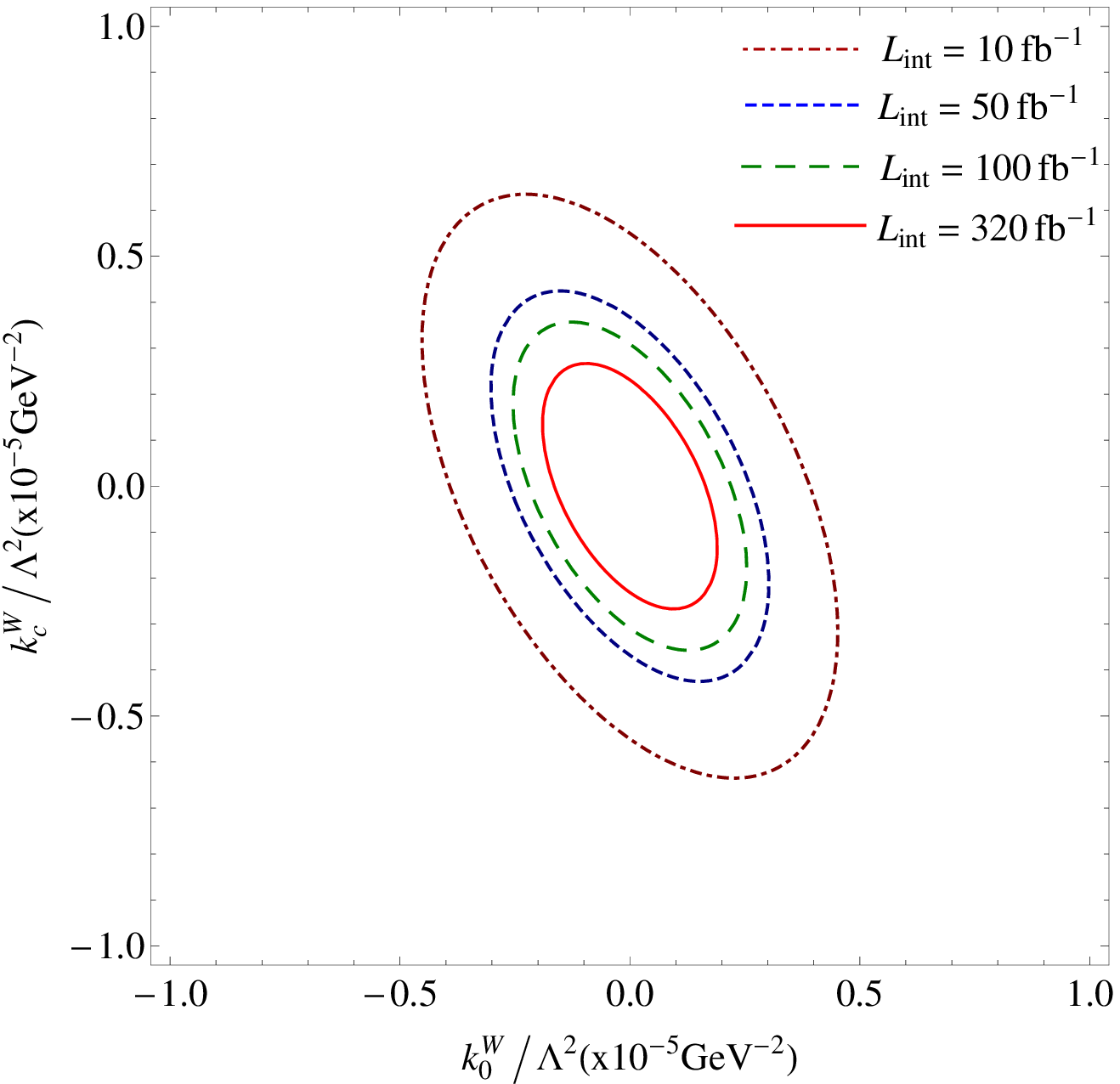}
\caption{The same as Fig. 25 but for $\sqrt{s}=1.5$ TeV.
\label{fig26}}
\end{figure}

\begin{figure}
\includegraphics[width=0.8\columnwidth]{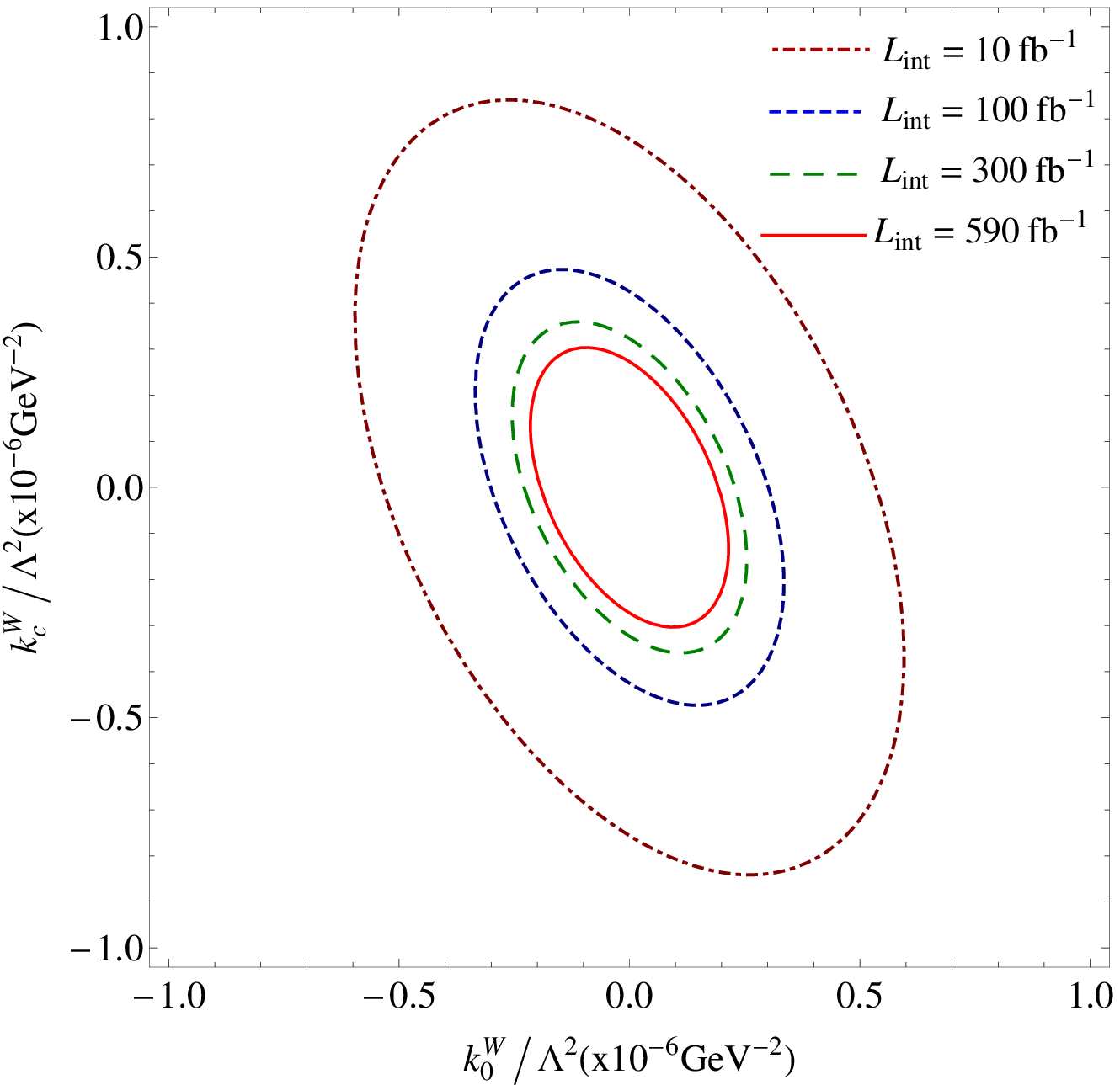}
\caption{The same as Fig. 25 but for $\sqrt{s}=3$ TeV.
\label{fig27}}
\end{figure}

In Figs. \ref{fig22}-\ref{fig24}, we present $95\%$ C.L. contours
for anomalous $\frac{k_{0}^{W}}{\Lambda^{2}}$ and
$\frac{k_{c}^{W}}{\Lambda^{2}}$ couplings for the process
$e^{+}e^{-}\rightarrow e^{+}\gamma^{*} \gamma^{*}e^{-}\rightarrow
e^{+}W^{+} W^{-} Z e^{-}$ at the CLIC for various integrated
luminosities and center-of-mass energies. As we can see from Fig.
\ref{fig24}, the best sensitivities on
$\frac{k_{0}^{W}}{\Lambda^{2}}$ and $\frac{k_{c}^{W}}{\Lambda^{2}}$
through this process are [$-1.38\times10^{-6}$, $1.38\times10^{-6}$]
and [$-1.96\times10^{-6}$, $1.96\times10^{-6}$], respectively for
$L_{int}=590$ fb$^{-1}$ at the CLIC. Also, the same contours for the
process $\gamma \gamma\rightarrow W^{+} W^{-}Z $ are given in Figs.
\ref{fig25}-\ref{fig27}. From two-parameter contours in
Fig.\ref{fig27}, the sensitivities for
$\frac{k_{0}^{W}}{\Lambda^{2}}$ and $\frac{k_{c}^{W}}{\Lambda^{2}}$
are obtained as [$-2.15\times10^{-6}$, $2.15\times10^{-6}$ ] and
[$-3.03\times10^{-6}$, $3.03\times10^{-6}$ ].

We can compare the obtained sensitivities on anomalous couplings by
using statistical significance
\begin{eqnarray}
SS=\frac{|\sigma_{tot}-\sigma_{SM}|}{\sqrt{\sigma_{SM}}} L_{int}
\end{eqnarray}
by assuming $\sqrt{s}=3$ TeV with the integrated luminosity of 590
fb$^{-1}$. Once again, we take into account leptonic decay channel
of the final state $Z$ boson and hadronic decay channel of W boson
for two processes. We obtain 3 (5) $\sigma$ observation
sensitivities on the anomalous couplings from the
$e^{+}e^{-}\rightarrow e^{+}\gamma^{*} \gamma^{*}e^{-}\rightarrow
e^{+}W^{+} W^{-} Z e^{-}$  process;
\begin{eqnarray*}
  &&-1.35(-1.74)\times10^{-6} < \frac{k_{0}^{W}}{\Lambda^{2}}< 1.35 (1.74)\times10^{-6} \\
 &&-1.90(-2.45)\times10^{-6} < \frac{k_{c}^{W}}{\Lambda^{2}}< 1.90 (2.45)\times10^{-6} \\
  &&-1.46(-1.89)\times10^{-6} < \frac{k_{2}^{m}}{\Lambda^{2}}< 1.46 (1.89)\times10^{-6} \\
   &&-1.29(-1.66)\times10^{-6} < \frac{a_{n}}{\Lambda^{2}}< 1.29 (1.66)\times10^{-6} \\
\end{eqnarray*}
and from the $\gamma \gamma\rightarrow W^{+} W^{-}Z$ process;

\begin{eqnarray*}
&&-2.14 (-2.76)\times10^{-7} < \frac{k_{0}^{W}}{\Lambda^{2}}< 2.14 (2.76)\times10^{-7}   \\
 &&-3.02 (-3.90)\times10^{-7} < \frac{k_{c}^{W}}{\Lambda^{2}}< 3.02 (3.90)\times10^{-7} \\
  &&-2.33 (-3.01)\times10^{-7} < \frac{k_{2}^{m}}{\Lambda^{2}}< 2.33 (3.01)\times10^{-7} \\
   &&-2.09 (-2.71)\times10^{-7} < \frac{a_{n}}{\Lambda^{2}}< 2.09 (2.71)\times10^{-7}. \\
\end{eqnarray*}
The obtained sensitivities using signal significance at 5 $\sigma$
are approximately 1.5 times better than the best sensitivities
obtained from $\chi^2$ analysis at 95\% C.L..
\section{Conclusions}

The linear $e^{-}e^{+}$ colliders will provide an important
opportunity to probe $e \gamma$ and $\gamma \gamma$ collisions at
high energies. In $e \gamma$ and $\gamma \gamma$ collisions, high
energy real photons can be obtained by converting the incoming
lepton beams into photon beams via the Compton backscattering
mechanism. In addition, high-energy accelerated $e^{-}$ and $e^{+}$
beams at the linear colliders radiate quasi-real photons, and thus
$e \gamma^{*}$ and $\gamma^{*} \gamma^{*}$ collisions are produced
from the $e^{-}e^{+}$ process itself. Therefore, $e \gamma^{*}$ and
$\gamma^{*} \gamma^{*}$ collisions at these colliders can occur
spontaneously apart from $e \gamma$ and $\gamma \gamma$ collisions.
In the literature, Refs. \cite{lhc1,lin2} only examined the
sensitivities on $\frac{a_{n}}{\Lambda^{2}}$ couplings through the
process $\gamma\gamma \rightarrow W^{+} W^{-}Z$ at future linear
colliders. As stated in Ref. \cite{lin3}, the $\gamma \gamma$
collisions can examine the sensitivities on
$\frac{a_{n}}{\Lambda^{2}}$ with a higher precision with respect to
the $e \gamma$ and $e^{-}e^{+}$ collisions. For this reason, we
compare our sensitivities with the results of Ref. \cite{mur}. For
$\frac{a_{n}}{\Lambda^{2}}$ couplings, $\gamma \gamma$ collisions at
the 3 TeV CLIC with an integrated luminosity of $ 590$ fb$^{-1}$
enable us to improve the sensitivities by almost a factor of five
with respect to sensitivities coming from $e^{-}e^{+}$ collisions.
Also, our sensitivities show that $\gamma \gamma$ collisions provide
anomalous $\frac{k_{2}^{m}}{\Lambda^{2}}$ couplings with a better
than the $e^{-}e^{+}$ collisions. On the other hand, we can see that
the sensitivities on $\frac{k_{0}^{W}}{\Lambda^{2}}$ and
$\frac{k_{c}^{W}}{\Lambda^{2}}$ expected to be obtained for the
future $\gamma \gamma$ colliders with $\sqrt{s}=3$ TeV are roughly 2
times worse than the sensitivities in Ref. \cite{mur}. We find that
the sensitivities obtained for four different
$\frac{k_{0}^{W}}{\Lambda^{2}}$, $\frac{k_{c}^{W}}{\Lambda^{2}}$ and
$\frac{k_{2}^{m}}{\Lambda^{2}}$ and $\frac{a_{n}}{\Lambda^{2}}$
couplings from the process $\gamma\gamma \rightarrow W^{+} W^{-}Z$
are approximately an order of magnitude more restrictive with
respect to the main process $e^{+}e^{-} \rightarrow e^{+}\gamma^{*}
\gamma^{*} e^{-} \rightarrow e^{+} W^{+} W^{-} Z e^{-}$ which is
obtained by integrating the cross section for the subprocess
$\gamma^{*} \gamma^{*}\rightarrow W^{+} W^{-}Z$ over the effective
photon luminosity. The process $\gamma\gamma (\gamma^{*}
\gamma^{*})\rightarrow W^{+} W^{-}Z$ includes only interactions
between the gauge bosons, causing more apparent possible deviations
from the expected value of SM \cite{lin3}. Therefore, in this paper,
we analyze the CP-conserving parameters
$\frac{k_{0}^{W}}{\Lambda^{2}}$, $\frac{k_{c}^{W}}{\Lambda^{2}}$ and
$\frac{k_{2}^{m}}{\Lambda^{2}}$ and CP-violating parameter
$\frac{a_{n}}{\Lambda^{2}}$ on the anomalous quartic $WWZ\gamma$
gauge couplings through the processes $\gamma \gamma \rightarrow
W^{+} W^{-}Z $ obtained by laser-backscattering distributions and
$e^{+}e^{-} \rightarrow e^{+}\gamma^{*} \gamma^{*} e^{-} \rightarrow
e^{+} W^{+} W^{-} Z e^{-}$ derived by EPA distributions at the CLIC.
The $\gamma \gamma$ collisions seem to be the best place to test
$\frac{k_{2}^{m}}{\Lambda^{2}}$ and $\frac{a_{n}}{\Lambda^{2}}$
which are the anomalous quartic couplings involving photons.
Therefore, the CLIC as photon-photon collider provides an ideal
platform to examine anomalous quartic $WWZ\gamma$ gauge couplings at
high energies.

\appendix*
\section{The anomalous vertex functions derived from CP-violating and CP-conserving effective Lagrangians}

The anomalous $W^{+} (p_{1}^{\alpha}) W^{-}(p_{2}^{\beta}) Z (k_{2}^{\nu}) \gamma (k_{1}^{\mu})$ vertex function obtained from CP-violating effective $\textit{L}_{n}$ Lagrangian is given below
\begin{eqnarray}
&&i\frac{\pi\alpha}{4 \textmd{cos}\,\, \theta_{W}\Lambda^{2}}a_{n}[g_{\alpha\nu}[g_{\beta\mu}\, k_{1}.(k_{2}-p_{1})-k_{1\beta}.(k_{2}-p_{1})_{\mu}] \nonumber \\
&&-g_{\beta\nu}[g_{\alpha\mu}\, k_{1}.(k_{2}-p_{2})-k_{1\alpha}.(k_{2}-p_{2})_{\mu}] \nonumber \\ &&+g_{\alpha\beta}[g_{\nu\mu}k_{1}.(p_{1}-p_{2})-k_{1\nu}.(p_{1}-p_{2})_{\mu}] \nonumber \\
&&-k_{2\alpha}(g_{\beta\mu}k_{1\nu}-g_{\nu\mu}k_{1\beta})+k_{2\beta}(g_{\alpha\mu}k_{1\nu}-g_{\nu\mu}k_{1\alpha}) \nonumber \\
&&-p_{2\nu}(g_{\alpha\mu}k_{1\beta}-g_{\beta\mu}k_{1\alpha})+p_{1\nu}(g_{\beta\mu}k_{1\alpha}-g_{\alpha\mu}k_{1\beta}) \nonumber \\
&&+p_{1\beta}(g_{\nu\mu}k_{1\alpha}-g_{\alpha\mu}k_{1\nu})+p_{2\alpha}(g_{\nu\mu}k_{1\beta}-g_{\beta\mu}k_{1\nu})].
\end{eqnarray}
The anomalous $W^{+} (p_{1}^{\alpha}) W^{-}(p_{2}^{\beta}) Z
(k_{2}^{\nu}) \gamma (k_{1}^{\mu})$ vertex functions obtained from
CP-conserving effective $\textit{W}_{0}^{Z}, \textit{W}_{c}^{Z},
\textit{W}_{1}^{Z}, \textit{W}_{2}^{Z}$ and $\textit{W}_{3}^{Z}$ can
be written as follows, respectively
\begin{eqnarray}
2ie^{2}g^{2}g_{\alpha\beta}[g_{\mu\nu}(k_{1}.k_{2})-k_{1\nu}k_{2\mu}],
\end{eqnarray}

\begin{eqnarray}
&&i\frac{e^{2}g^{2}}{2}[(g_{\mu\alpha}g_{\nu\beta}+g_{\nu\alpha}g_{\mu\beta})(k_{1}.k_{2})+g_{\mu\nu}(k_{2\beta}k_{1\alpha}+k_{1\beta}k_{2\alpha}) \nonumber \\
&&-k_{2\mu}k_{1\alpha}g_{\nu\beta}-k_{2\beta}k_{1\nu}g_{\mu\alpha}-k_{2\alpha}k_{1\nu}g_{\mu\beta}-k_{2\mu}k_{1\beta}g_{\nu\alpha}],
\end{eqnarray}

\begin{eqnarray}
ieg_{z}g^{2}((g_{\mu \alpha}k_{1}.p_{1}-p_{1\mu}k_{1\alpha})g_{\nu
\beta}+(g_{\mu \beta}k_{1}.p_{2} - p_{2\mu}k_{1\beta})g_{\nu
\alpha})
\end{eqnarray}

\begin{eqnarray}
&&i\frac{eg_{z}g^{2}}{2}((k_{1}.p_{1}+k_{1}.p_{2})g_{\mu \nu}g_{\alpha \beta}-(k_{1 \alpha}p_{1\beta}+k_{1\beta}p_{2\alpha})g_{\mu \nu} \nonumber \\
&&-(p_{1\mu}+p_{2\mu})k_{1\nu}g_{\alpha \beta}+(p_{1\beta}g_{\mu\alpha}+p_{2\alpha}g_{\mu \beta})k_{1\nu}),
\end{eqnarray}

\begin{eqnarray}
&&i\frac{eg_{z}g^{2}}{2}(k_{1}.p_{1}g_{\mu \beta}g_{\nu \alpha}+k_{1}.p_{2}g_{\mu \alpha}g_{\nu \beta}+(p_{1\nu}-p_{2\nu})k_{1\beta} g_{\mu \alpha} \nonumber \\
&& -(p_{1\nu}-p_{2\nu})k_{1\alpha} g_{\mu \beta}-p_{1\mu}k_{1\beta}g_{\nu \alpha}-p_{2\mu}k_{1\alpha}g_{\nu \beta}).
\end{eqnarray}

\pagebreak

\pagebreak

\begin{table}
\caption{The sensitivities of the anomalous
$\frac{k_{0}^{W}}{\Lambda^{2}}$ and $\frac{k_{c}^{W}}{\Lambda^{2}}$
couplings through the process $\gamma \gamma\rightarrow W^{+} W^{-}Z
$ at the CLIC with $\sqrt{s}=0.5,1.5$ and $3$ TeV for various
integrated luminosities. \label{tab1}}
\begin{ruledtabular}
\begin{tabular} {cccc}
$\sqrt{s}$ (TeV)& $L_{int}$(fb$^{-1}$)& $\frac{k_{0}^{W}}{\Lambda^{2}}$(GeV$^{-2}$)& $\frac{k_{c}^{W}}{\Lambda^{2}}$ (GeV$^{-2}$)\\
\hline
$0.5$& $10$& $[-9.39; 8.91]\times 10^{-5}$& $[-1.36;\, 1.34]\times 10^{-4}$ \\

$0.5$& $50$& $[-6.36;\, 5.88]\times 10^{-5}$& $[-9.14;\, 8.93]\times 10^{-5}$  \\

$0.5$& $100$& $[-5.39;\, 4.91]\times 10^{-5}$& $[-7.69;\, 7.49]\times 10^{-5}$\\

$0.5$& $230$& $[-4.42;\, 3.94]\times 10^{-5}$& $[-6.27;\, 6.06]\times 10^{-5}$ \\
\hline
$1.5$& $10$& $[-3.50;\, 3.50]\times 10^{-6}$& $[-4.92;\, 4.92]\times 10^{-6}$ \\

$1.5$& $50$& $[-2.34;\, 2.34]\times 10^{-6}$& $[-3.29;\, 3.29]\times 10^{-6}$  \\

$1.5$& $100$& $[-1.97;\, 1.97]\times 10^{-6}$& $[-2.77;\, 2.77]\times 10^{-6}$\\

$1.5$& $320$& $[-1.47;\, 1.47]\times 10^{-6}$& $[-2.07;\, 2.07]\times 10^{-6}$ \\
\hline
$3$& $10$& $[-4.80;\, 4.80]\times 10^{-7}$& $[-6.77;\, 6.77]\times 10^{-7}$ \\

$3$& $100$&$[-2.69;\, 2.69]\times 10^{-7}$& $[-3.81;\, 3.81]\times 10^{-7}$  \\

$3$& $300$& $[-2.05;\, 2.05]\times 10^{-7}$& $[-2.89;\, 2.89]\times 10^{-7}$\\

$3$& $590$& $[-1.73;\, 1.73]\times 10^{-7}$& $[-2.44;\, 2.44]\times 10^{-7}$ \\
\end{tabular}
\end{ruledtabular}
\end{table}

\begin{table}
\caption{The sensitivities of the anomalous
$\frac{k_{2}^{m}}{\Lambda^{2}}$ and $\frac{a_{n}}{\Lambda^{2}}$
couplings through the process $\gamma \gamma\rightarrow W^{+} W^{-}Z
$ at the CLIC with $\sqrt{s}=0.5,1.5$ and $3$ TeV for various
integrated luminosities. \label{tab2}}
\begin{ruledtabular}
\begin{tabular} {cccc}
$\sqrt{s}$ (TeV)& $L_{int}$(fb$^{-1}$)& $\frac{k_{2}^{m}}{\Lambda^{2}}$(GeV$^{-2}$)& $\frac{a_{n}}{\Lambda^{2}}$ (GeV$^{-2}$)\\
\hline
$0.5$& $10$& $[-2.17; 2.09]\times 10^{-4}$& $[-3.59;\, 3.59]\times 10^{-4}$ \\

$0.5$& $50$& $[-1.47;\, 1.39]\times 10^{-4}$& $[-2.40;\, 2.40]\times 10^{-4}$  \\

$0.5$& $100$& $[-1.24;\, 1.16]\times 10^{-4}$& $[-2.02;\, 2.02]\times 10^{-4}$\\

$0.5$& $230$& $[-1.01;\, 0.94]\times 10^{-5}$& $[-1.64;\, 1.64]\times 10^{-4}$ \\
\hline
$1.5$& $10$& $[-5.22;\, 5.21]\times 10^{-6}$& $[-6.04;\, 6.04]\times 10^{-6}$ \\

$1.5$& $50$& $[-3.49;\, 3.49]\times 10^{-6}$& $[-4.04;\, 4.04]\times 10^{-6}$  \\

$1.5$& $100$& $[-2.93;\, 2.93]\times 10^{-6}$& $[-3.39;\, 339]\times 10^{-6}$\\

$1.5$& $320$& $[-2.19;\, 2.19]\times 10^{-6}$& $[-2.54;\, 2.54]\times 10^{-6}$ \\
\hline
$3$& $10$& $[-5.23;\, 5.23]\times 10^{-7}$& $[-4.70;\, 4.70]\times 10^{-7}$ \\

$3$& $100$&$[-2.94;\, 2.94]\times 10^{-7}$& $[-2.64;\, 2.64]\times 10^{-7}$  \\

$3$& $300$& $[-2.23;\, 2.23]\times 10^{-7}$& $[-2.01;\, 2.01]\times 10^{-7}$\\

$3$& $590$& $[-1.89;\, 1.89]\times 10^{-7}$& $[-1.74;\, 1.74]\times 10^{-7}$ \\
\end{tabular}
\end{ruledtabular}
\end{table}

\begin{table}
\caption{The sensitivities of the anomalous
$\frac{k_{0}^{W}}{\Lambda^{2}}$ and $\frac{k_{c}^{W}}{\Lambda^{2}}$
couplings through the process $e^{+}e^{-}\rightarrow e^{+}\gamma^{*}
\gamma^{*}e^{-}\rightarrow e^{+}W^{+} W^{-} Z e^{-}$ at the CLIC
with $\sqrt{s}=1.5$ and $3$ TeV for various integrated luminosities.
\label{tab3}}
\begin{ruledtabular}
\begin{tabular} {cccc}
$\sqrt{s}$ (TeV)& $L_{int}$(fb$^{-1}$)& $\frac{k_{0}^{W}}{\Lambda^{2}}$(GeV$^{-2}$)& $\frac{k_{c}^{W}}{\Lambda^{2}}$ (GeV$^{-2}$)\\
\hline
$0.5$& $10$& $[-3.72;\, 3.71]\times 10^{-4}$& $[-5.48;\, 5.41]\times 10^{-4}$ \\

$0.5$& $50$& $[-2.49;\, 2.48]\times 10^{-4}$& $[-3.68;\, 3.61]\times 10^{-4}$  \\

$0.5$& $100$& $[-2.09;\, 2.08]\times 10^{-4}$& $[-3.10;\, 3.02]\times 10^{-4}$\\

$0.5$& $230$& $[-1.70;\, 1.69]\times 10^{-4}$& $[-2.52;\, 2.45]\times 10^{-4}$ \\
\hline
$1.5$& $10$& $[-1.83;\, 1.83]\times 10^{-5}$& $[-2.58;\, 2.58]\times 10^{-5}$ \\

$1.5$& $50$& $[-1.23;\, 1.23]\times 10^{-5}$& $[-1.72;\, 1.72]\times 10^{-5}$  \\

$1.5$& $100$& $[-1.03;\, 1.03]\times 10^{-5}$& $[-1.45;\, 1.45]\times 10^{-5}$\\

$1.5$& $320$& $[-7.71;\, 7.71]\times 10^{-6}$& $[-1.08;\, 1.08]\times 10^{-5}$ \\
\hline
$3$& $10$& $[-3.03;\, 3.03]\times 10^{-6}$& $[-4.27;\, 4.27]\times 10^{-6}$ \\

$3$& $100$&$[-1.70;\, 1.70]\times 10^{-6}$& $[-2.40;\, 2.40]\times 10^{-6}$  \\

$3$& $300$& $[-1.29;\, 1.29]\times 10^{-6}$& $[-1.82;\, 1.82]\times 10^{-6}$\\

$3$& $590$& $[-1.09;\, 1.09]\times 10^{-6}$& $[-1.54;\, 1.54]\times 10^{-6}$ \\
\end{tabular}
\end{ruledtabular}
\end{table}

\begin{table}
\caption{The sensitivities of the anomalous
$\frac{k_{2}^{m}}{\Lambda^{2}}$ and $\frac{a_{n}}{\Lambda^{2}}$
couplings through the process $e^{+}e^{-}\rightarrow e^{+}\gamma^{*}
\gamma^{*}e^{-}\rightarrow e^{+}W^{+} W^{-} Z e^{-}$ at the CLIC
with $\sqrt{s}=1.5$ and $3$ TeV for various integrated luminosities.
\label{tab4}}
\begin{ruledtabular}
\begin{tabular} {cccc}
$\sqrt{s}$ (TeV)& $L_{int}$(fb$^{-1}$)& $\frac{k_{2}^{m}}{\Lambda^{2}}$(GeV$^{-2}$)& $\frac{a_{n}}{\Lambda^{2}}$ (GeV$^{-2}$)\\
\hline
$0.5$& $10$& $[-8.46;\, 8.20]\times 10^{-4}$& $[-1.35;\, 1.35]\times 10^{-4}$ \\

$0.5$& $50$& $[-5.70;\, 5.44]\times 10^{-4}$& $[-9.02;\, 9.02]\times 10^{-5}$  \\

$0.5$& $100$& $[-4.81;\, 4.55]\times 10^{-4}$& $[-7.56;\, 7.56]\times 10^{-5}$\\

$0.5$& $230$& $[-3.93;\, 3.67]\times 10^{-4}$& $[-6.15;\, 6.15]\times 10^{-5}$ \\
\hline
$1.5$& $10$& $[-2.72;\, 2.72]\times 10^{-5}$& $[-3.10;\, 3.10]\times 10^{-5}$ \\

$1.5$& $50$& $[-1.82;\, 1.82]\times 10^{-5}$& $[-2.07;\, 2.07]\times 10^{-5}$  \\

$1.5$& $100$& $[-1.53;\, 1.53]\times 10^{-5}$& $[-1.74;\, 1.74]\times 10^{-5}$\\

$1.5$& $320$& $[-1.14;\, 1.14]\times 10^{-5}$& $[-1.30;\, 1.30]\times 10^{-5}$ \\
\hline
$3$& $10$& $[-3.28;\, 3.28]\times 10^{-6}$& $[-2.89;\, 2.89]\times 10^{-6}$ \\

$3$& $100$&$[-1.84;\, 1.84]\times 10^{-6}$& $[-1.62;\, 1.62]\times 10^{-6}$  \\

$3$& $300$& $[-1.40;\, 1.40]\times 10^{-6}$& $[-1.24;\, 1.24]\times 10^{-6}$\\

$3$& $590$& $[-1.18;\, 1.18]\times 10^{-6}$& $[-1.04;\, 1.04]\times 10^{-6}$ \\
\end{tabular}
\end{ruledtabular}
\end{table}
\end{document}